\documentclass[11pt]{article}
\usepackage{graphicx,amssymb,amstext,amsmath,amssymb,graphicx,bm,dsfont,gensymb}
\usepackage[top=30mm, bottom=30mm, left=35mm, right=35mm]{geometry}
\usepackage{natbib}
\usepackage{xr}
\allowdisplaybreaks
\begin{document}

\begin{center}
{\bf\Large Conditional Modelling of Spatio-Temporal Extremes for Red Sea Surface Temperatures}\\
{\large Emma S.\ Simpson and Jennifer L.\ Wadsworth\\}
Lancaster University
\end{center}

\begin{abstract}
Recent extreme value theory literature has seen significant emphasis on the modelling of spatial extremes, with comparatively little consideration of spatio-temporal extensions. This neglects an important feature of extreme events: their evolution over time. Many existing models for the spatial case are limited by the number of locations they can handle; this impedes extension to space-time settings, where models for higher dimensions are required. Moreover, the spatio-temporal models that do exist are restrictive in terms of the range of extremal dependence types they can capture. Recently, conditional approaches for studying multivariate and spatial extremes have been proposed, which enjoy benefits in terms of computational efficiency and an ability to capture both asymptotic dependence and asymptotic independence. We extend this class of models to a spatio-temporal setting, conditioning on the occurrence of an extreme value at a single space-time location. We adopt a composite likelihood approach for inference, which combines information from full likelihoods across multiple space-time conditioning locations. We apply our model to Red Sea surface temperatures, show that it fits well using a range of diagnostic plots, and demonstrate how it can be used to assess the risk of coral bleaching attributed to high water temperatures over consecutive days.
\end{abstract}


\section{Introduction}\label{sec:intro}
There are many situations where modelling extreme events may be of interest; understanding such events can allow us to prepare for, and potentially mitigate, their effect. For environmental extremes, such as high temperatures or intense rainfall, the spatial extent of an extreme event is of particular concern, with possible focus on determining regions that may be affected by a given phenomenon. Moreover, situations may be exacerbated if the extreme events persist for a period of time, indicating that temporal modelling also requires attention. To understand and prepare for such spatio-temporal extreme events, we require modelling techniques motivated by extreme value theory. In this paper, we focus on modelling surface temperature extremes in the Red Sea by extending the so-called conditional approaches of \cite{Heffernan2004}, \cite{Heffernan2007} and \cite{Wadsworth2019} to a spatio-temporal setting. We condition on observations at a single space-time location being above some high threshold, and construct a model for other locations across a spatio-temporal domain. 

The tail dependence properties of variables are an important consideration in multivariate extreme value modelling. Many models are limited in terms of whether they can capture asymptotic dependence, when variables can take their extreme values simultaneously, or asymptotic independence, when the largest extremes occur separately. More formally, consider random variables $X(s,t)$ and $X(s+h_\textsc{s},t+h_\textsc{t})$ at two space-time locations separated by spatial distance $h_\textsc{s}$ and temporal lag $h_\textsc{t}$, with $X(\cdot,\cdot)\sim F_{\cdot,\cdot}$. To assess the extremal dependence between these variables, one option is to consider the conditional survivor function
\begin{align}
\chi(u;h_\textsc{s},h_\textsc{t}) = \Pr\left[F_{s+h_\textsc{s},t+h_\textsc{t}}\{X(s+h_\textsc{s},t+h_\textsc{t})\}>u\mid F_{s,t}\{X(s,t)\}>u\right],
\label{eqn:chiu}\end{align}
for some value of $u$ close to 1. A common measure of asymptotic dependence is obtained by the relation $\chi(h_\textsc{s},h_\textsc{t})=\lim_{u\rightarrow 1}\chi(u;h_\textsc{s},h_\textsc{t})$, with $\chi(h_\textsc{s},h_\textsc{t})=0$ corresponding to asymptotic independence, and $\chi(h_\textsc{s},h_\textsc{t})\in(0,1]$ revealing asymptotic dependence between $X(s,t)$ and $X(s+h_\textsc{s},t+h_\textsc{t})$. In practice, we expect that this dependence decreases as $h_\textsc{s}$ and $h_\textsc{t}$ increase.

Existing methods for modelling spatial extremes include max-stable processes (see \cite{Davison2012}), with spatio-temporal extensions given by \cite{Davis2013} and \cite{Huser2014}. These processes arise as asymptotically-justified extensions of the generalised extreme value distribution for modelling univariate extremes \citep{vonMises1936, Jenkinson1955} when pointwise maxima are taken over spatio-temporal domains. A drawback of max-stable processes is that they are computationally intensive to fit, limiting the number of space-time locations one can feasibly handle, though the recently proposed semiparametric approach of \cite{Buhl2019} aims to address this point. A further potential drawback of this class of models is their ability to capture only asymptotic dependence. If max-stable models are fitted to asymptotically independent data, results will be too conservative as we extrapolate into the tail. They therefore lend themselves to modelling extremes only in small spatio-temporal domains, where asymptotic dependence is more likely to arise. It is also unclear which data applications would be of interest for temporally dependent maxima, since the process of taking block maxima usually leads to approximately independent observations, meaning space-time max-stable process models may not be so useful in practice. To address some of these issues, \cite{Morris2017} propose a modelling technique that allows for different tail dependence features across space. In particular, they suggest partitioning the spatial locations to allow for asymptotic dependence for nearby sites, and asymptotic independence between those that are further apart, modelled via a skew-$t$ process. The temporal dependence in their model is captured by an AR(1) process, but in many cases it may be desirable to have models that are not restricted to one type of temporal dependence, allowing for both asymptotic dependence and asymptotic independence in time.

Recently, there has been increasing interest in using conditional approaches, first introduced by \cite{Heffernan2004}, to model extreme events. In the multivariate setting, this involves conditioning on a single variable being above some high threshold, and modelling the behaviour of the remaining variables. In the spatial case \citep{Wadsworth2019}, additional structural properties are exploited to construct models conditioning on extreme values at a single spatial location. Conditional extremes models have the advantage of being able to capture a range of tail dependence behaviours, while being computationally efficient compared to other methods. These appealing features lead us to extend the spatial conditional extremes methodology to a spatio-temporal setting, which is particularly helpful for handling the additional burden that comes from expanding the domain from space to space-time. Important new considerations for this spatio-temporal extension include: whether or not the model should exhibit separability in space and time; how to separate the data into individual events for inference; and the construction of model diagnostics that allow us to assess the fit across both domains.

\begin{figure}[t]
\begin{center}
\includegraphics[clip, trim=0cm 0.5cm 0cm 0.75cm, width=0.8\textwidth]{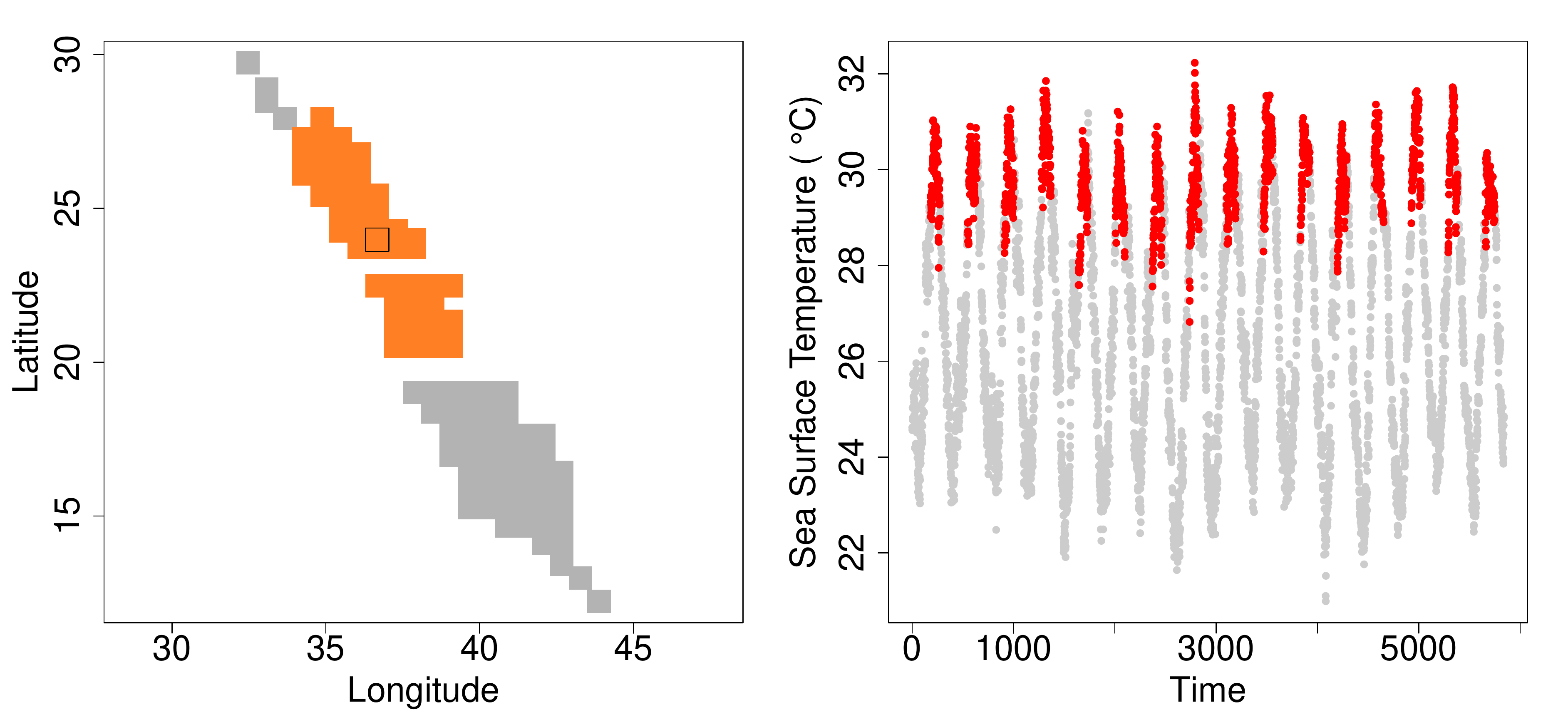}
\caption{Locations of observations, with the northerly region we study in orange (left), and sea surface temperature data for the location outlined in black (right). The red points correspond to the months July to September.}
\label{fig:dataLocations}
\end{center}
\end{figure}

We demonstrate our new spatio-temporal model through an application to Red Sea surface temperatures. This is particularly of interest as continual high sea temperatures can lead to problems for marine life, with coral bleaching being a notable concern; see \cite{McClanahan2007} and \cite{Huser2020}. While corals can recover from high sea temperatures that occur over a short period of time, prolonged extreme events can lead to irreparable damage and coral mortality, highlighting the importance of studying the temporal aspect of sea surface temperature extremes. A dataset relating to the Red Sea was the focus of the recent EVA conference data challenge \citep{Huser2020}. The challenge involved modelling temperature anomalies rather than the raw surface temperatures we study here, with the aim being to develop methods for missing data, which is also not our focus. Approaches developed as a result of this challenge include \cite{CastroCamilo2020} and \cite{Vlah2020}.

To study surface temperature extremes in the Red Sea, we have access to 16 years of daily sea surface temperature data, for the 108 locations shown in the left panel of Figure~\ref{fig:dataLocations}. These locations are a subset of $0.05\degree\times 0.05\degree$ grid cells covering the whole of the Red Sea. The data are obtained from a combination of satellite measurements and in situ readings; see \cite{Donlon2012}. Our initial investigations indicated differing behaviour in the north and south of the Red Sea, as well as in the most north-westerly locations, corresponding to the Gulf of Suez; this issue is also identified by \cite{Huser2020}. Since the model we propose assumes spatio-temporal stationarity, we focus on the northerly region highlighted in orange, containing 54 locations, and comment further on approaches to deal with spatial non-stationarity in Section~\ref{sec:discussion}. To remove the need to account for seasonality in the data, we focus only on the warmest months of July to September. Example data for one location are shown in the right panel of Figure~\ref{fig:dataLocations}, with other locations in the northerly region exhibiting similar seasonal behaviour. Focussing on this time period is consistent with the claims of \cite{Maynard2008}, that coral bleaching is most likely during summer months. 

The remainder of the paper is constructed as follows. We detail our proposed model in Section~\ref{sec:modelling}, and discuss inference and simulation in Section~\ref{sec:inference}. In Section~\ref{sec:diagnostics}, we present a variety of diagnostic tools for assessing model fit, as well as results on Red Sea surface temperatures in the context of coral bleaching. Section~\ref{sec:discussion} concludes with a discussion of potential extensions.

\section{Modelling}\label{sec:modelling}
\subsection{Model assumption}\label{sec:assumption}
Let $w=(s,t)$ denote a space-time location, and suppose we are interested in the stationary spatio-temporal process $\{X(w): w\in\mathcal{S}\times\mathcal{T}\}$, with $\mathcal{S}$ and $\mathcal{T}$ denoting the space and time domains, respectively. As with all conditional extremes approaches, we require that each marginal distribution of the process has a standard exponential upper tail, i.e., $\Pr\{X(w)>x\}\sim e^{-x}$, as $x\rightarrow\infty$, for each $w\in\mathcal{S}\times\mathcal{T}$. In practice, this can be achieved via a transformation. We focus on Laplace margins, as suggested by \cite{Keef2013}, which we obtain by applying the probability integral transform via a rank transformation to each spatial location.

Conditioning on the process at a single space-time location, $w_0\in\mathcal{S}\times\mathcal{T}$, being above a high threshold $u$, we assume that for a finite number of space-time locations $w_1,\dots,w_{dm}$, corresponding to $d$ spatial locations at $m$ time-points, there exist functions $a_{w-w_0}(\cdot)$ and $b_{w-w_0}(\cdot)$ such that
\begin{align}
	\Bigg(\left[\frac{X(w_i) - a_{w_i-w_0}\left\{X(w_0)\right\}}{b_{w_i-w_0}\left\{X(w_0)\right\}}\right]_{i=1,\dots,dm},X(w_0)-u\Bigg)&\Bigg\vert X(w_0)>u\nonumber\\ 
	&\hspace{-1cm}\xrightarrow{d} \bigg(\left\{Z^0(w_i)\right\}_{i=1,\dots,dm}, E\bigg),
\label{eqn:mainAssumption}
\end{align}
as $u\rightarrow\infty$. That is, after normalisation, the process $\{X(w):w\in\mathcal{S}\times\mathcal{T}\}$ converges to $\{Z^0(w):w\in\mathcal{S}\times\mathcal{T}\}$ in the sense of finite dimensional distributions. The variable $E$ is independent of $\{Z^0(w)\}$ in the limit and follows a standard exponential distribution. The function $a_{w-w_0}(x)$ should take values in $[0,x]$, with $a_0(x)=x$, and be non-increasing with respect to $\|s-s_0\|$ and $\|t-t_0\|$. This implies that the process $\{Z^0(w)\}$, subsequently referred to as the residual process, must satisfy $Z^0(w_0)=0$. We note that although we are working with finite dimensional distributions here, these are given useful structure by having the spatio-temporal index domain.

To model $\{X(w)\}$ using assumption~\eqref{eqn:mainAssumption}, choices need to be made about the form of the normalising functions and the residual process; we discuss these in Sections~\ref{subsec:normalisingFunctions}~and~\ref{subsec:residualProcess}, respectively. \cite{Wadsworth2019} demonstrate that~\eqref{eqn:mainAssumption} holds for a wide variety of underlying spatial dependence structures, detailing specific forms of $a_{s-s_0}(\cdot)$, $b_{s-s_0}(\cdot)$ and $\{Z^0(s):s\in\mathcal{S}\}$. It will therefore hold for analogous dependence structures with spatio-temporal indexing. We check the validity of assumption~\eqref{eqn:mainAssumption} for our Red Sea data in Section~\ref{sec:diagnostics}.

\subsection{The normalising functions $a_{w-w_0}(\cdot)$ and $b_{w-w_0}(\cdot)$}\label{subsec:normalisingFunctions}
In order to exploit assumption~\eqref{eqn:mainAssumption} for modelling purposes, we impose parametric forms on the normalising functions $a_{w-w_0}(\cdot)$ and $b_{w-w_0}(\cdot)$. In the spatial case, under the assumption of isotropy, \cite{Wadsworth2019} suggest taking $a_{s-s_0}(x)=x\alpha_\textsc{s}(s-s_0)$, with 
\begin{align}
\alpha_\textsc{s}(s-s_0)= 
  \begin{cases}
    1, & \|s-s_0\|\leq\Delta_{\textsc{s}},\\
    \exp\left[-\left\{(\left\|s-s_0\|-\Delta_{\textsc{s}}\right)/\lambda_{\textsc{s}}\right\}^{\kappa_{\textsc{s}}}\right], & \|s-s_0\|>\Delta_{\textsc{s}},
  \end{cases}
\label{eqn:aFunction}
\end{align}
for $\|s-s_0\|$ denoting the distance between $s$ and $s_0$, $\lambda_{\textsc{s}}>0$, $\kappa_{\textsc{s}}\in[0,2]$ and $\Delta_{\textsc{s}}\geq 0$. If $\Delta_{\textsc{s}}>0$, this yields $a_{s-s_0}(x)=x$ for $\|s-s_0\|\leq\Delta_{\textsc{s}}$, allowing for asymptotic dependence up to distances of $\Delta_{\textsc{s}}$ from the conditioning site, and asymptotic independence beyond that. Although~\eqref{eqn:aFunction} assumes isotropy, anisotropy is straightforward to handle, as discussed in Section~\ref{subsec:anisotropy}. In the spatio-temporal setting, a possible extension of this is to take
\begin{align}
a_{w-w_0}(x) = x\alpha_\textsc{s}(s-s_0)\alpha_\textsc{t}(t-t_0),
\label{eqn:separable_a}
\end{align}
with $\alpha_\textsc{t}(t-t_0)$ defined analogously to~\eqref{eqn:aFunction}, where we recall that the conditioning site is denoted by $w_0=(s_0,t_0)$, and more generally we set $w=(s,t)$. This allows for asymptotic dependence within some space-time neighbourhood of the conditioning site, controlled by the parameters $(\Delta_{\textsc{s}},\Delta_{\textsc{t}})$, and asymptotic independence outside this neighbourhood. In modelling the Red Sea surface temperatures, we found the inclusion of these parameters not to have a significant effect; we therefore fix $\Delta_{\textsc{s}},\Delta_{\textsc{t}}=0$, corresponding to an asymptotically independent model.

Normalising function~\eqref{eqn:separable_a} exhibits space-time separability, but one could instead consider a non-separable form for $a_{w-w_0}(\cdot)$. A natural choice is to exploit the class of known, non-separable covariance functions, such as those introduced by \cite{Cressie1999} and studied further by \cite{Gneiting2002}. One possibility, based on the class of \cite{Gneiting2002}, is to take
\begin{align}
	a_{w-w_0}(x) = x\left(\lambda_{\textsc{t}}\|t-t_0\|^{2\kappa_{\textsc{t}}}+1\right)^{-1}\exp\left\{-\frac{\lambda_{\textsc{s}}\|s-s_0\|^{2\kappa_{\textsc{s}}}}{\left(\lambda_{\textsc{t}}\|t-t_0\|^{2\kappa_{\textsc{t}}}+1\right)^{\eta\kappa_{\textsc{s}}}}\right\},
\label{eqn:nonsepAlpha}
\end{align}
for $\lambda_{\textsc{s}}, \lambda_{\textsc{t}}>0$, $\kappa_{\textsc{s}},\kappa_{\textsc{t}}\in(0,1]$ and $\eta\in[0,1]$. The parameter $\eta$ controls the strength of the interaction between space and time, with $\eta=0$ corresponding to separability. We note that while covariance functions provide a convenient way to introduce non-separability into the function $a_{w-w_0}(x)$, as well as allowing us to satisfy the conditions that $a_{w-w_0}(x)\in[0,x]$ and is non-increasing with respect to $\|s-s_0\|$ and $\|t-t_0\|$, we are not restricted to this class of models since positive definiteness is not required. Asymptotic dependence could be incorporated into \eqref{eqn:nonsepAlpha} via an approach analogous to the use of $(\Delta_{\textsc{s}},\Delta_{\textsc{t}})$ in the separable case.

Three theoretically-motivated forms for the normalising function $b_{s-s_0}(\cdot)$ are proposed by \cite{Wadsworth2019}. We focus on one such example in the spatio-temporal extension, taking 
\begin{align}
	b_{w-w_0}(x) = 1 + \left\{a_{w-w_0}(x)\right\}^\beta,
\label{eqn:bFunction}
\end{align}
for some $\beta\in[0,1]$. This can be used for either of our proposed forms of $a_{w-w_0}(x)$. The advantage of using~\eqref{eqn:bFunction} as the model for $b_{w-w_0}(x)$ is that the function can vary with distance, while being controlled by only a single extra parameter, but other options could also be used.

\subsection{The residual process $\{Z^0(w)\}$}\label{subsec:residualProcess}
We construct a model for the residual process by first considering the stationary space-time Gaussian process $\left\{Z(w):w\in\mathcal{S}\times\mathcal{T}\right\}$, with mean $\mu$ and standard deviation $\sigma>0$. We propose using a separable covariance function with powered exponential components in space and time, i.e., the covariance corresponding to sites $w_i=(s_i,t_i)$ and $w_j=(s_j,t_j)$ is 
\begin{align}
	\sigma^2 \cdot \exp\left\{-\left(\|s_i-s_j\|/\phi_{\textsc{s}}\right)^{p_{\textsc{s}}} \right\} \cdot \exp\left\{-\left(\|t_i-t_j\|/\phi_{\textsc{t}}\right)^{p_{\textsc{t}}} \right\}.
\label{eqn:separableCovariance}
\end{align}
To ensure the condition that $Z^0(w_0)=0$ is satisfied, the residual process is taken as the conditional Gaussian process $\{Z(w)\}\vert Z(w_0)=0$. We note that taking this Gaussian form for the residual process does not induce Gaussianity in the process $\{X(w)\}$ due to the appearance of the normalising functions.

Recall that we are interested in modelling observations at space-time locations $w_1,\dots,w_{dm}$. The corresponding mean vector of the Gaussian process $\{Z(w):w\in\mathcal{S}\times\mathcal{T}\}$ is $\bm{\mu}=\mu\cdot\bm{1}_{dm}\in\mathbb{R}^{dm}$, with $\bm{1}_{dm}$ denoting a vector of 1s, and we denote the covariance matrix, constructed via function~\eqref{eqn:separableCovariance}, as $\Sigma\in\mathbb{R}^{dm\times dm}$. These two components can be partitioned to represent the space-time locations to be modelled, and the conditioning site, i.e., $\bm{\mu}=\mu\cdot\left(\bm{1}_{dm-1},1\right)$, and \[
\Sigma=\begin{bmatrix} \Sigma^* & \Sigma_0^* \\ \Sigma_0^{*T} & \sigma^2 \end{bmatrix}, \text{ for } \Sigma^*\in\mathbb{R}^{(dm-1)\times (dm-1)} \text{ and } \Sigma_0^*\in\mathbb{R}^{dm-1}.
\]
Then the conditional Gaussian process $\{Z(w)\}\vert Z(w_0)=0$ follows a multivariate Gaussian distribution with mean $\bm{\mu}_{\vert 0}\in\mathbb{R}^{dm-1}$ and covariance matrix $\Sigma_{\vert 0}\in\mathbb{R}^{(dm-1)\times(dm-1)}$ of the form
\begin{align}
\bm{\mu}_{\vert 0} = \mu\cdot\left(\bm{1}_{dm-1} - \sigma^{-2}\Sigma_0^*\right);~~~\Sigma_{\vert 0} = \Sigma^* - \sigma^{-2}\Sigma_0^*\Sigma_0^{*T}.
\label{eqn:conditionalMeanVar}
\end{align}

In place of~\eqref{eqn:separableCovariance}, we could also consider a non-separable covariance function. However, most of the structure in our model is captured by the normalising functions $a_{w-w_0}(x)$ and $b_{w-w_0}(x)$, and separability in the covariance function of $\{Z(w)\}$ actually induces non-separability in $\{Z^0(w)\}$. We therefore choose to focus only on covariance function~\eqref{eqn:separableCovariance}.

\cite{Wadsworth2019} propose what they term a delta-Laplace distribution for the marginal form of their residual process $\{Z^0(s):s\in\mathcal{S}\}$, which has univariate Gaussian and Laplace distributions as special cases. We expect independence between observations at large spatial lags, and the delta-Laplace distribution allows for the recovery of the original Laplace margins in such cases. It is possible to include such a marginal distribution in our spatio-temporal extension, however, we found that at the distances studied in our Red Sea example, this is not necessary, and retaining the Gaussian margins of the conditional Gaussian process leads to improvements in terms of computational efficiency. 

\subsection{Spatial anisotropy}\label{subsec:anisotropy}
We account for spatial anisotropy through a transformation of the spatial coordinates, setting
\begin{align}
s^*=\begin{pmatrix}
1 & 0 \\
0 & 1/L
\end{pmatrix}
\begin{pmatrix}
\cos\theta & -\sin\theta \\
\sin\theta & \cos\theta
\end{pmatrix}
s,
\label{eqn:anisotropy}
\end{align}
where the parameter $\theta\in[-\pi/2,0]$ controls rotation, and $L>0$ relates to the amount by which the coordinates are stretched, with $L=1$ recovering the isotropic model. Such an approach is often termed \emph{geometric anisotropy}. These parameters are estimated, along with the other model parameters, using the methods discussed in Section~\ref{subsec:compositeLikelihood}.


\section{Inference and simulation}\label{sec:inference}
\subsection{Extreme events for inference}\label{subsec:chi}
We propose carrying out inference using a likelihood approach. To reduce the computational burden of these calculations, we first aim to separate the data into shorter space-time blocks that allow us to capture the evolution of single extreme events.

Suppose that an extreme value at a particular location is defined as any observation above the high threshold $u$. One option, proposed by \cite{HuserWadsworth2019}, is to consider clusters of days with extreme values at any site, separated by $k$ days with no extreme values, to belong to a single extreme event, for some value $k$. This provides an intuitive way to separate extreme events, but for our data, we found this often produced long clusters that were not computationally feasible to handle. A simpler alternative is to divide the data into blocks of $m$ days, for some choice of $m$, taking care not to have blocks that span multiple years. If the chosen block-length, $m$, does not exactly divide the length of each summer period, we remove an appropriate number of observations at the end of each one. The value of $m$ should be large enough to provide sufficient information about temporal changes in the extremal dependence, but small enough to ensure computational feasibility.

To determine an appropriate value of $m$, we investigate the spatio-temporal dependence present in our data, using the measure $\chi(u;h_\textsc{s},h_\textsc{t})$ defined in~\eqref{eqn:chiu}. For a finite number of realisations of $X(s,t)$ and $X(s+h_\textsc{s},t+h_\textsc{t})$, it is not possible to evaluate $\chi(h_\textsc{s},h_\textsc{t})=\lim_{u\rightarrow 1}\chi(u;h_\textsc{s},h_\textsc{t})$ itself, so we instead consider an empirical estimate of \eqref{eqn:chiu}, denoted by $\hat\chi(u;h_\textsc{s},h_\textsc{t})$. That is, for $n$ pairs of observations $\left(x_1(s,t),x_1(s+h_\textsc{s},t+h_\textsc{t})\right),\dots,$ $\left(x_n(s,t),x_n(s+h_\textsc{s},t+h_\textsc{t})\right)$,
\[
\hat\chi(u;h_\textsc{s},h_\textsc{t})=\frac{\sum\limits_{i=1}^n\mathds{1}\{x_i(s+h_\textsc{s},t+h_\textsc{t})>q_u,x_i(s,t)>q_u\}}{\sum\limits_{i=1}^n\mathds{1}\{x_i(s,t)>q_u\}},
\]
with $q_u$ denoting the quantile of the Laplace distribution corresponding to threshold $u$. We evaluate $\hat\chi(u;h_\textsc{s},h_\textsc{t})$ for observations from each pair of spatial locations, and at increasing time lags. The results are shown by the grey points in Figures~\ref{fig:chi_cluster5_95}~and~\ref{fig:chi_cluster5_975} for $u=0.95$ and $u=0.975$, respectively; these plots will later be used to assess model fit. As expected, the strongest extremal dependence occurs for pairs of nearby locations and at short time lags. From these plots, it appears that the pairwise spatial dependence becomes stable for time lags greater than three or four. We therefore expect that taking $m\geq 4$ should enable us to sufficiently capture the changing spatio-temporal dependence in our process. For the remainder of the paper we take $m=5$, corresponding to a maximum time lag of four, but results for other values of $m$ are presented in Section~\ref{sup:differentClusterLength} of the Supplementary Material.  

\subsection{Inference}\label{subsec:compositeLikelihood}
To estimate the parameters of our model, we adopt a composite likelihood approach similar to that of \cite{Wadsworth2019}, where the idea is to allow different space-time locations to act as the conditioning site, and pool information from each one. In our case, these conditioning sites correspond to all combinations of the spatial locations in Figure~\ref{fig:dataLocations}, and the time points in our chosen blocks of length $m$. With $w^*=(s^*,t)$ denoting locations under transformation~\eqref{eqn:anisotropy}, the likelihood contribution of conditioning site $w_i$ is
\begin{align*}
\mathcal{L}_i(\bm\theta) = \prod_{\ell=1}^{n_i} \Bigg(f^{Z^0}_{i}&\left[\left\{\frac{x_j^\ell - a_{w^*_j-w^*_i}(x_i^\ell)}{b_{w^*_j-w^*_i}(x_i^\ell)}\right\}_{j\in\{1,\dots,dm\}\setminus\{i\}}\right] \cdot\prod_{j\in\{1,\dots,dm\}\setminus\{i\}}b_{w^*_j-w^*_i}(x_i^\ell)^{-1}\Bigg),
\end{align*}
for the $n_i$ observations where $x(w^*_i)>u$, where $x^\ell_{j}$ denotes the value of the $\ell$th such observation at space-time location $w^*_j$, and $f^{Z^0}_{i}$ represents the density of the residual process $\{Z^0(w^*)\}$ at all locations except $w^*_i$. Here, $\bm\theta$ represents all model parameters relating to the residual process $Z^0$ and normalising functions $a_{w-w_0}$ and $b_{w-w_0}$, and is discussed more explicitly for the models we use in Section~\ref{subsec:Zdiagnostic}.

The overall likelihood is given by $\mathcal{L}(\bm\theta)=\prod_{i=1}^{dm}\mathcal{L}_{i}(\bm\theta)$. This is a composite likelihood since the same observation can be included in the likelihood calculation multiple times, depending on how many space-time locations take values above the threshold $u$. The parameters $\bm\theta$ can be estimated using standard maximum likelihood techniques, and we take $u$ to be the 0.95 quantile of the marginal Laplace distributions, which results in an average of 73 threshold exceedances per spatial location, or approximately 14 per space-time location. We use \texttt{optim} in \texttt{R} for inference, which works well in general, but can require repeated iterations to ensure convergence is reached. We subsequently drop the $s^*$ and $w^*$ notation, but note that geometric anisotropy is included for the remainder of the paper.

\subsection{Simulation given extremes at a single location}\label{subsec:simulation}
Many of the diagnostics we introduce in Section~\ref{sec:diagnostics} and in the Supplementary Material rely on the simulation of events from our fitted models. To simulate a single spatio-temporal event $\{x(w):w\in\mathcal{S}\times\mathcal{T}\}$, conditioning on $X(w_0)>v$, for $v\geq u$, i.e., at a level at least as extreme as where we fit the model, the steps are:
\begin{enumerate}
\item simulate $e\sim\rm{Exp}(1)$, and let $x(w_0)=v+e$;
\item simulate $\{z^0(w):w\in\mathcal{S}\times\mathcal{T}\}$ from the model for the residual process defined in Section~\ref{subsec:residualProcess};
\item set $\{x(w):w\in\mathcal{S}\times\mathcal{T}\} = a_{w-w_0}\left[x(w_0)\right] + b_{w-w_0}\left[x(w_0)\right]\cdot\{z^0(w)\}$.
\end{enumerate}

We note that the resulting simulations are on the Laplace scale, which is sufficient in our setting as we are particularly interested in the dependence structure of our variables, and in Section~\ref{subsec:corals} we will use quantiles to approximate critical levels associated with coral bleaching. One could marginally invert the rank transform to obtain simulations on the original scale.

\subsection{Importance sampling}\label{subsec:importance}
Rather than conditioning on extreme values at a single space-time location, it may be more useful to condition on situations where extremes occur anywhere in some spatio-temporal domain. \cite{Wadsworth2019} introduce an importance sampling algorithm to handle this case, which is straightforward to adapt to our spatio-temporal setting. Suppose we are interested in conditioning on $\max_{w\in D}X(w) >v$, for some $v\geq u$, where $D$ denotes a subset of locations in the spatio-temporal domain $\mathcal{S}\times\mathcal{T}$, which may or may not be identical to the locations $\{w_1,\dots,w_{dm}\}$ used for inference. For some function $g(\cdot)$, the aim is to estimate a quantity of the form
\begin{align}
\mathbb{E}\left[g\left\{X(w):w\in D\right\}\Big\vert \max\limits_{w\in D}X(w) >v\right].
\label{eqn:importanceSample}
\end{align}
This is achieved by taking the following steps:
\begin{enumerate}
\item sample a location $\tilde{w}$ from the set $D$, each with probability $|D|^{-1}$;
\item simulate a single observation using the method in Section~\ref{subsec:simulation}, with $\tilde{w}$ as the conditioning site;
\item repeat steps 1 and 2 $n$ times, generating samples $\bm{X}_{D,1},\dots,\bm{X}_{D,n}$;
\item obtain an estimate of~\eqref{eqn:importanceSample} via
\[
\frac{\sum\limits_{j=1}^n \left[g(\bm{X}_{D,j})\cdot \left\{\sum\limits_{x\in \bm{X}_{D,j}}\mathds{1}(x>v)\right\}^{-1}\right]}{\sum\limits_{j=1}^n \left\{\sum\limits_{x\in \bm{X}_{D,j}}\mathds{1}(x>v)\right\}^{-1}}.
\]
\end{enumerate}
We demonstrate results using this approach in Section~\ref{subsec:ISresults}.


\section{Model diagnostics and results}\label{sec:diagnostics}
\subsection{Pairwise $\chi(u;h_\textsc{s},h_\textsc{t})$ in space and time}\label{subsec:chiSpaceTime}

\begin{figure}[!htbp]
\begin{center}
\includegraphics[width=0.8\textwidth]{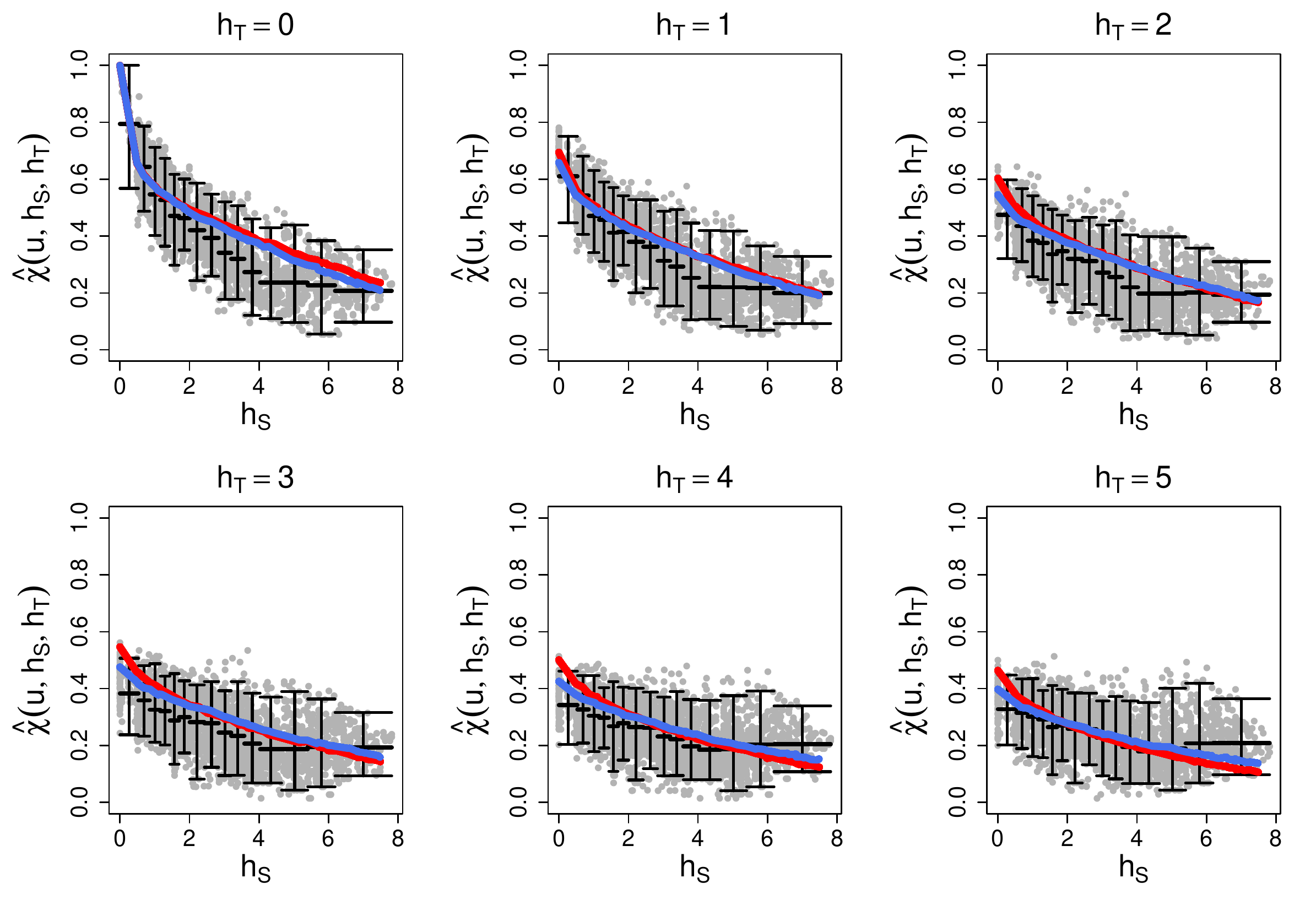}
\caption{Pairwise estimates of $\chi(0.95;h_\textsc{s},h_\textsc{t})$ for the Red Sea data (grey), with the mean and 0.025 and 0.975 quantiles for sections of the distances (black). The solid lines show $\chi(0.95;h_\textsc{s},h_\textsc{t})$ estimates for our fitted models: separable (red); non-separable (blue).}
\label{fig:chi_cluster5_95}\vspace{0.2cm}
\includegraphics[width=0.8\textwidth]{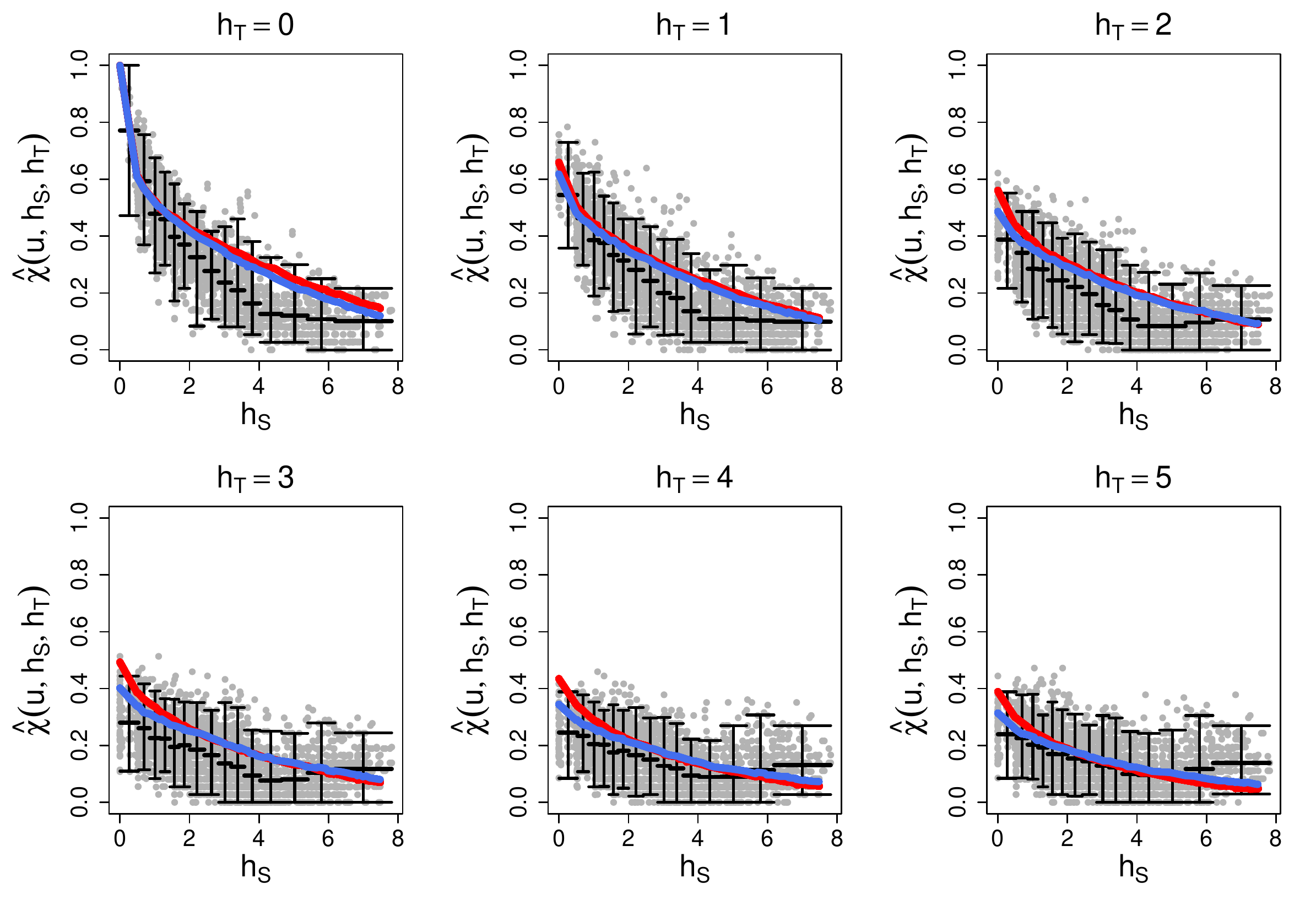}
\caption{Pairwise estimates of $\chi(0.975;h_\textsc{s},h_\textsc{t})$ for the Red Sea data (grey), with the mean and 0.025 and 0.975 quantiles for sections of the distances (black). The solid lines show $\chi(0.975;h_\textsc{s},h_\textsc{t})$ estimates for our fitted models: separable (red); non-separable (blue).}
\label{fig:chi_cluster5_975}
\end{center}
\end{figure}

As a first assessment of how well our model captures the spatio-temporal dependence present in the data, we again consider the measure $\chi(u;h_\textsc{s},h_\textsc{t})$ discussed in Section~\ref{subsec:chi}. In Figures~\ref{fig:chi_cluster5_95}~and~\ref{fig:chi_cluster5_975}, we compare estimates of $\chi(u;h_\textsc{s},h_\textsc{t})$ calculated from the data, to estimates obtained from 10,000 simulations from our fitted model, with both the separable model~\eqref{eqn:separable_a} and non-separable model~\eqref{eqn:nonsepAlpha} for $a_{w-w_0}(x)$, and $u=0.95,0.975$. The spatial distances are calculated using the Euclidean distance on locations under transformation~\eqref{eqn:anisotropy}. We demonstrate this transformation in Section~\ref{sup:anisotropy} of the Supplementary Material.

We see from Figures~\ref{fig:chi_cluster5_95}~and~\ref{fig:chi_cluster5_975} that the two models perform similarly well overall. The main difference is the performance at short spatial distances as the time lag increases, where the separable model overestimates the extremal dependence, and the non-separable model performs better. Due to this advantage, we use the non-separable model for the remainder of the paper.

\subsection{Diagnostic based on $\{Z^0(w)\}$}\label{subsec:Zdiagnostic}
After anisotropy parameters, the remaining parameters in our model can be separated into those related to the normalising functions $a_{w-w_0}(\cdot)$ and $b_{w-w_0}(\cdot)$, and those from the model of the residual process $\{Z^0(w)\}$. Under the non-separable model, the former is the set $\bm{\theta}_{\mbox{norm}}=\left(\lambda_{\textsc{s}},\kappa_{\textsc{s}},\lambda_{\textsc{t}},\kappa_{\textsc{t}},\eta,\beta\right)$ and the latter is $\bm{\theta}_{\mbox{res}}=\left(\mu,\sigma,\phi_{\textsc{s}},p_{\textsc{s}},\phi_{\textsc{t}},p_{\textsc{t}}\right)$. We obtain estimates of these parameters through maximising the composite likelihood discussed in Section~\ref{subsec:compositeLikelihood}, and denote these by $\hat{\bm{\theta}}_{\mbox{norm}}$ and $\hat{\bm{\theta}}_{\mbox{res}}$, respectively. 

Empirical realisations of the residual process can be obtained by normalising the observed data, transformed to Laplace margins, using functions~\eqref{eqn:nonsepAlpha}~and~\eqref{eqn:bFunction} with parameters $\hat{\bm{\theta}}_{\mbox{norm}}$, and conditioning on the observation at site $w_0$ being above the modelling threshold $u$. For this same choice of conditioning site, we can simulate directly from the multivariate Gaussian distribution with mean vector and covariance matrix in~\eqref{eqn:conditionalMeanVar} determined by parameter values $\hat{\bm{\theta}}_{\mbox{res}}$, in order to obtain realisations of our fitted residual process. If the model fits well, the empirical and fitted residuals should exhibit similar behaviour.

\begin{figure}[!htbp]
\begin{center}
\includegraphics[width=0.8\textwidth]{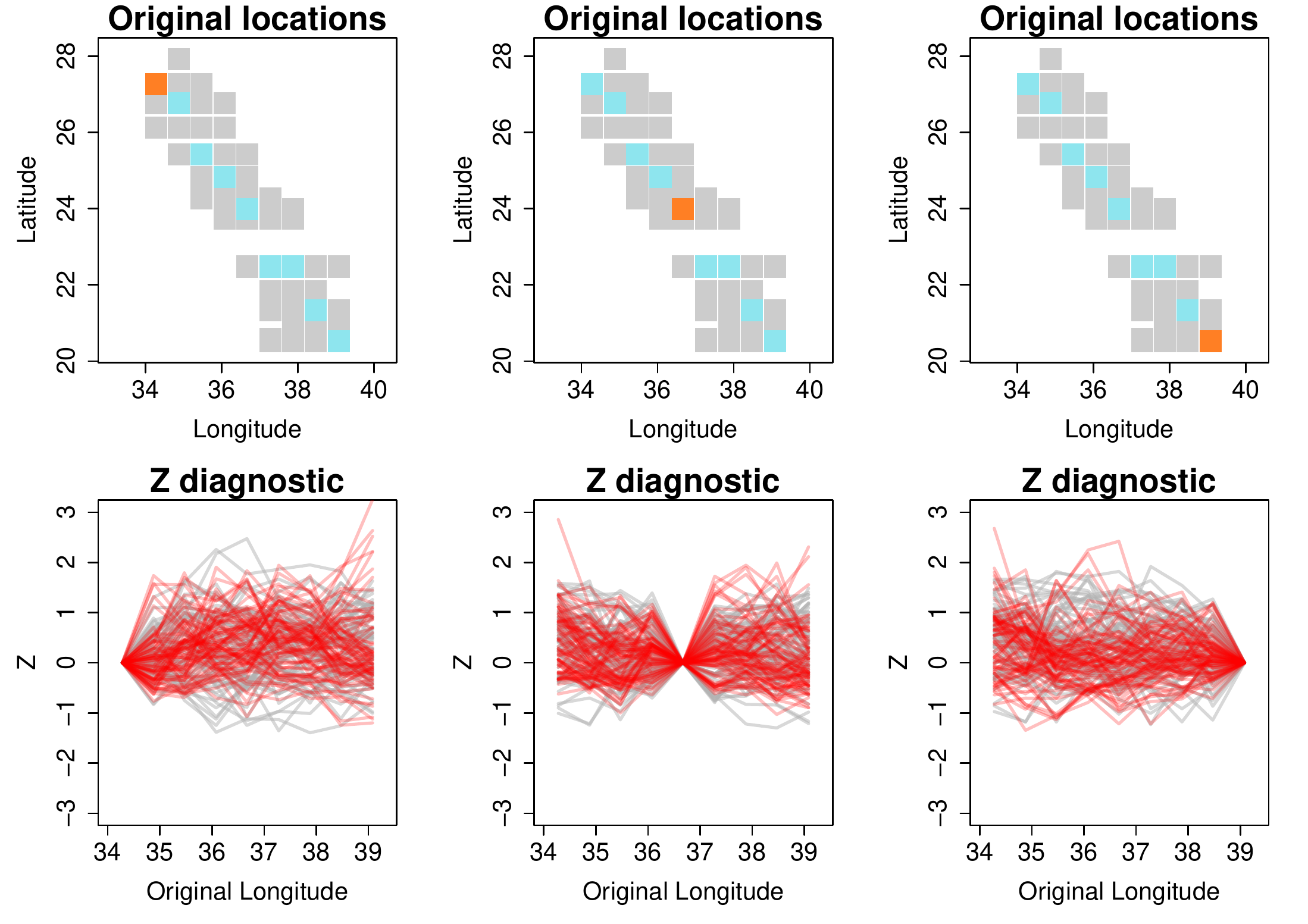}
\caption{Top row: locations of the conditioning sites (orange) and spatial transect (blue); bottom row: comparison of residuals simulated from the fitted model (grey) and empirical residuals (red) across space.}
\label{fig:spatialZdiag_cluster5}
\vspace{0.5cm}\includegraphics[width=0.8\textwidth]{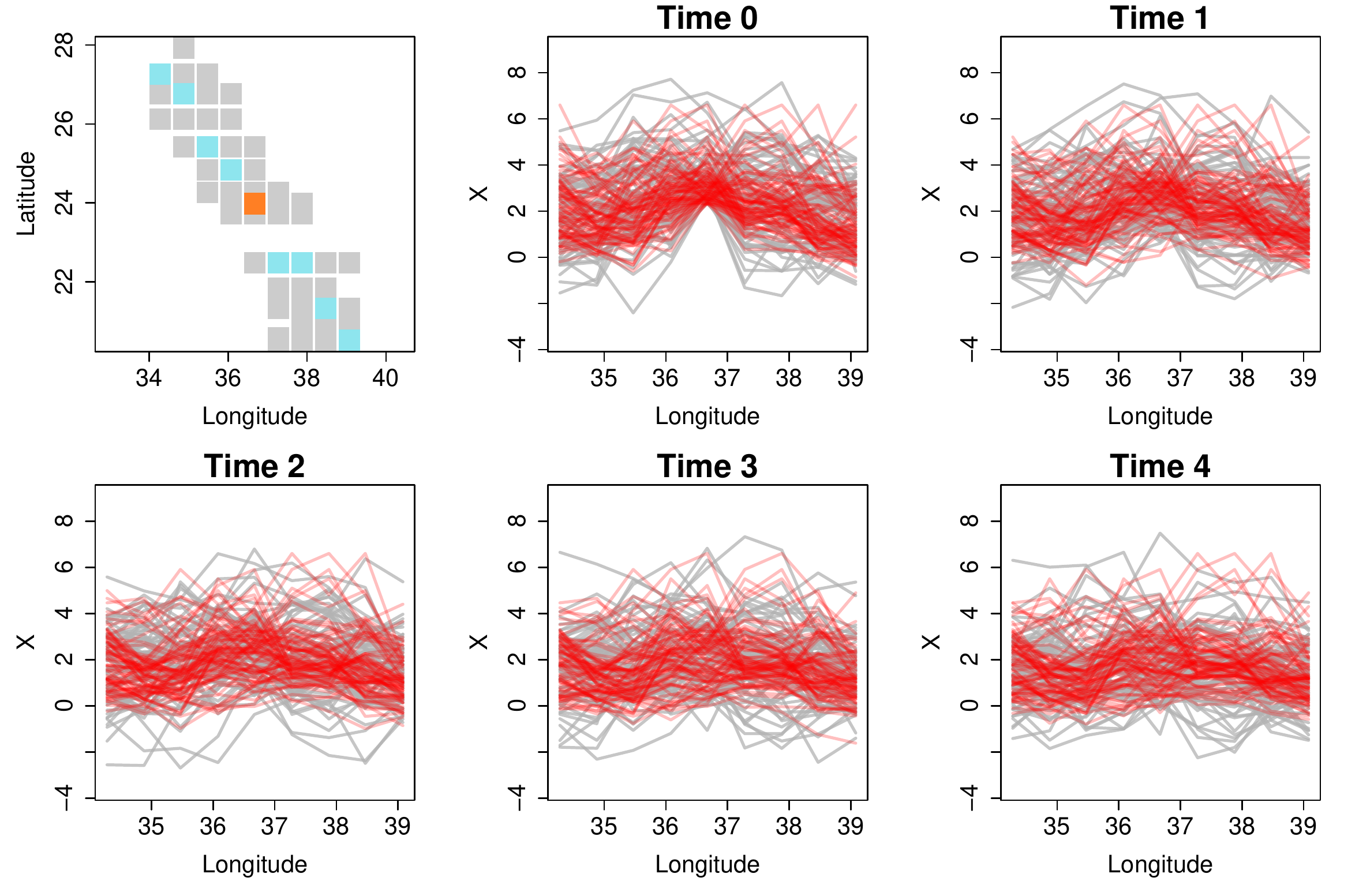}
\caption{Location of the conditioning site (orange) and spatial transect (blue), top left, and comparison of simulations from the fitted model (grey) and observed data (red) on Laplace scale at different time lags.}
\label{fig:Xdiag}
\end{center}
\end{figure}

For the model fitted using block-length $m=5$, Figure~\ref{fig:spatialZdiag_cluster5} demonstrates the behaviour of the empirical and fitted residuals across a diagonal spatial transect, for three different conditioning sites. This transect was chosen as it includes some of the longest spatial ranges present in our data. We take $u$ to be the 0.95 quantile of the observations in Laplace margins. The number of simulations from the fitted residual process was taken to be the same as the number of empirical observations. The diagnostic plots demonstrate reasonable agreement between the empirical and fitted residuals.

Figure~\ref{fig:spatialZdiag_cluster5} represents a purely spatial diagnostic. In Section~\ref{sup:Zdiagnostic} of the Supplementary Material, we present similar results to assess the temporal aspect of our model. These diagnostic plots also indicate a successful model fit.

\subsection{Diagnostic based on $\{X(w)\}$}
Another useful diagnostic is to consider simulations from our fitted model, obtained via the method outlined in Section~\ref{subsec:simulation}, conditioning on extreme values at some conditioning site $w_0=(s_0,0)$. We can then consider the propagation of these events forwards in time, and compare the simulated events to observations from the data on Laplace margins. In Figure~\ref{fig:Xdiag}, we demonstrate this technique for the same spatial transect as in Figure~\ref{fig:spatialZdiag_cluster5}, and for a single conditioning site $s_0$. In these plots, we also see reasonably similar behaviour between the simulated events and observed data.

\subsection{Evaluating parameter uncertainty}
Due to our use of a composite likelihood, it is necessary to evaluate the uncertainty in our parameter estimates, for which we use the bootstrap \citep{Efron1979}. For a time series of length $n$, \cite{Kunsch1989} proposes creating blocks of length $b$ containing consecutive observations, i.e., starting at observation $1, 2, \dots, n+1-b$. One can then randomly sample $n/b$ of these blocks, with replacement, to obtain each bootstrapped sample. We take this approach, but remove blocks that span more than one summer to ensure each block contains consecutive observations. Since we separate the data into blocks of length $m$ to carry out our likelihood inference, it is possible that not all the temporal dependence in the data has been captured in our model. We can account for this issue by taking a block-size $b>m$ for the bootstrapped samples. 

A drawback of the block bootstrap approach is that it can lead to non-stationarity in the bootstrapped samples, even if the original observations exhibit stationarity. \cite{Politis1994} propose a stationary bootstrapping approach to overcome this issue, where the first observation in each block is still sampled uniformly across all observations, but the lengths of the blocks are randomly sampled from a $Geom(p)$ distribution. The use of random block-lengths does not guarantee that the blocks, of fixed length $m$, that we subsequently create for our inferential approach contain consecutive observations, so we continue with the block bootstrap approach.

\begin{figure}
\begin{center}
\includegraphics[clip, trim=0cm 1.5cm 0cm 0cm, width=0.8\textwidth]{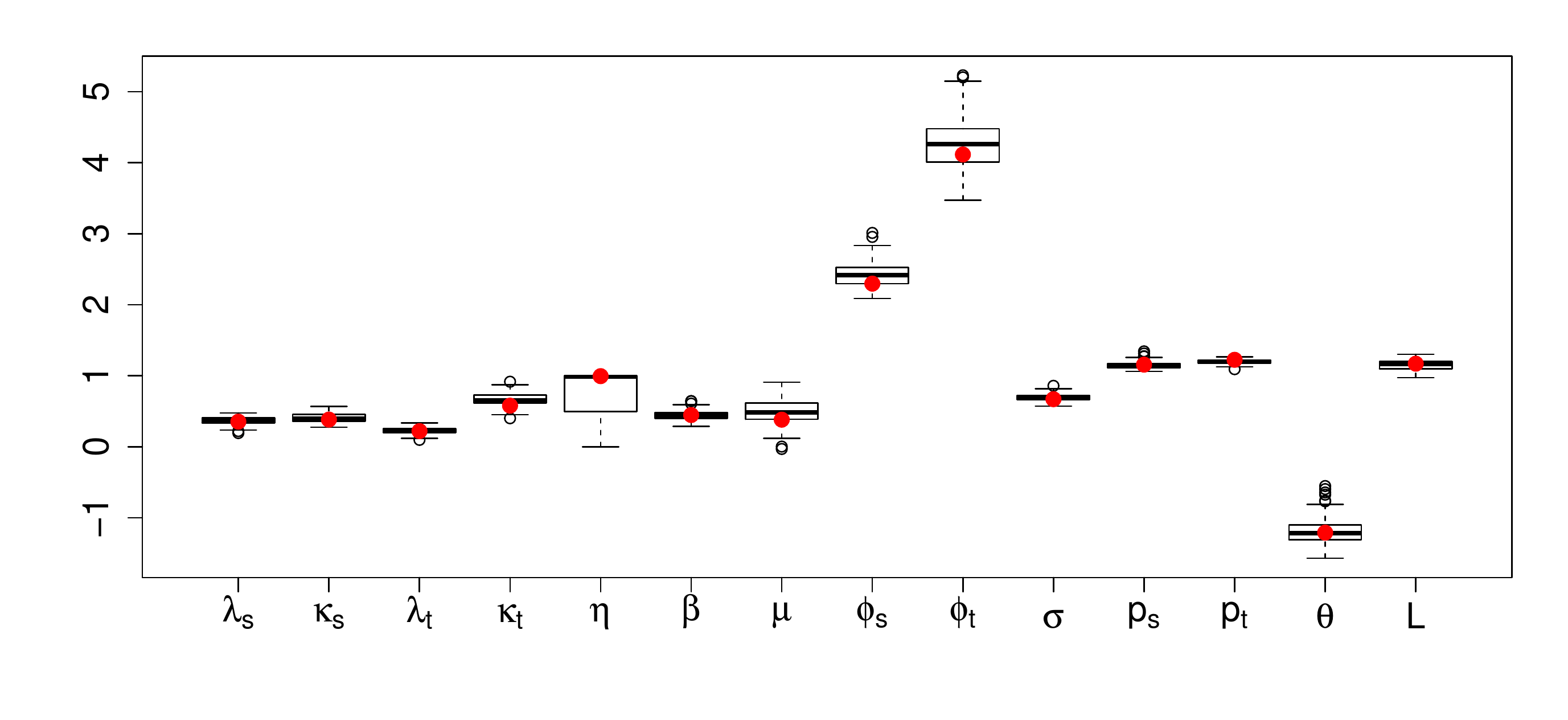}
\caption{Boxplots of bootstrapped parameter estimates for 100 bootstrapped samples, with the estimated parameters values for the original data shown in red.}
\label{fig:bootstrap}
\end{center}
\end{figure}

As in the preceding investigations, we fix the block-length used to carry out our inference as $m=5$. Then each summer contains 90 observations, and we have $n=90\times 16=1440$ observations overall. We take a block-length of $b=20$ for the bootstrap, with 72 blocks making up each of our bootstrapped samples. We fit the non-separable version of our model to 100 such samples, with boxplots of the resulting parameter estimates shown in Figure~\ref{fig:bootstrap}.

\subsection{Conditioning on an extreme at any location}\label{subsec:ISresults}
We now consider quantities of the form \eqref{eqn:importanceSample}, estimated via the importance sampling algorithm discussed in Section~\ref{subsec:importance}. This method can be applied for any set of space-time locations within the domain $\mathcal{S}\times\mathcal{T}$. In this section, we focus on the 54 locations where we have observations, at five time-points, corresponding to our chosen block-length $m$, i.e., a total of 270 locations $w_1,\dots,w_{270}$.

\begin{figure}
\begin{center}
\includegraphics[clip, trim=0cm 0.5cm 0cm 0.5cm, width=0.8\textwidth]{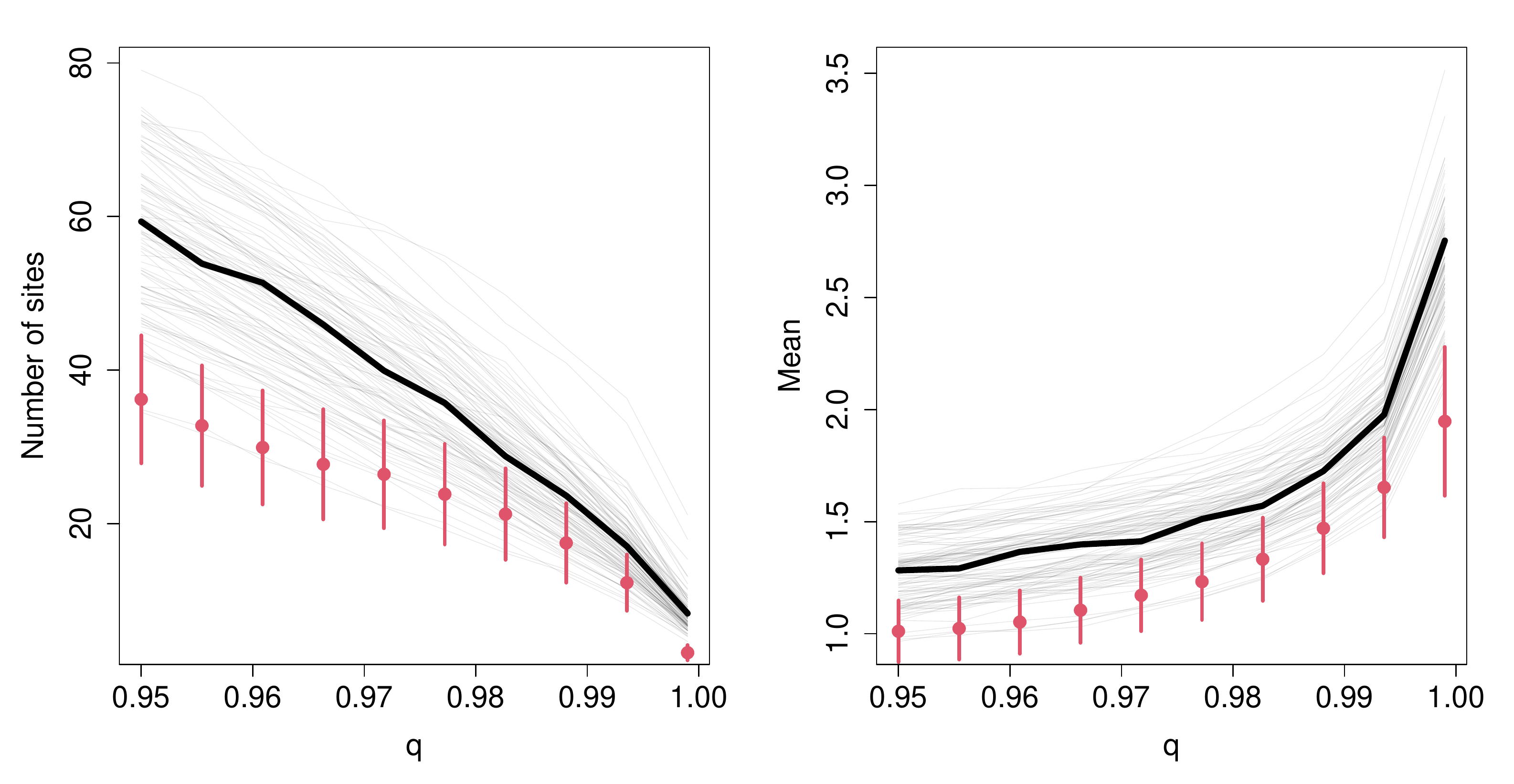}
\caption{Importance sampling results for the number of sites exceeding a threshold (left) and the average value of the temperatures on the Laplace scale (right), given an exceedance anywhere in the domain. The threshold $v$ is given in terms of quantiles of the Laplace distribution. The solid black lines show results using our parameter estimates, and the grey lines show the bootstrapped results. The red circles provide empirical estimates from the data with 95\% confidence intervals.}
\label{fig:importance}
\end{center}
\end{figure}

In Figure~\ref{fig:importance}, we consider results for two different functions $g$. The left panel corresponds to the expected number of space-time locations that exceed the threshold $v$, given that  $\max_{i=1,\dots,270}X(w_i)>v$. The right panel shows results based on the average temperature over all sites, on the Laplace scale, under the same conditioning event. We simulate $n=10,000$ importance samples for each threshold. To demonstrate uncertainty in the estimates, we present results obtained using the bootstrapped parameter estimates in Figure~\ref{fig:bootstrap}, and compare these to empirical estimates obtained using the data.

The results in Figure~\ref{fig:importance} reveal that our model overestimates the number of simultaneously extreme locations, particularly at lower thresholds, as well as the average temperature across all locations. These discrepancies may be explained by the results in Figures~\ref{fig:chi_cluster5_95}~and~\ref{fig:chi_cluster5_975}; for short time-lags, we slightly overestimate the pairwise extremal dependence, the effect of which could have been exacerbated when considering results for all locations simultaneously. However, the empirical estimates are within the range of our bootstrapped calculations in both cases, so taking uncertainty into account, our model appears to perform adequately.

\subsection{Investigation into the risk of coral bleaching}\label{subsec:corals}
Sea surface temperatures can be used as an indicator of potentially damaging conditions for coral life. The average sea surface temperature at different coral reefs varies depending on geographic location, with different varieties of coral adapting to their different habitats. As such, there is no single, high temperature threshold to which coral bleaching can be attributed. While \cite{Jokiel2004} collate information about temperature thresholds associated with this phenomenon for a range of locations, none of these are particularly close to the northern Red Sea. Since the risk of coral bleaching increases when sea surface temperatures significantly exceed the norm for a particular location and time of year, rather than using a fixed threshold to explain coral bleaching, \cite{Genevier2019} propose using high quantiles that vary with space (an approach adopted by \cite{Hazra2019}), and time. \cite{Goreau2005} suggest that coral bleaching may occur if temperatures persistently exceed a level corresponding to one degree above the average temperature in the warmest month, which for our data is August. In Section~\ref{sup:coralQuantiles} of the Supplementary Material, we show that this is well approximated by taking the 0.961 quantile of summer sea surface temperatures for each location we study, so this is the level on which we focus our study.

Coral reefs are located around much of the coast of the northern Red Sea, with those in the north-west generating particularly high revenue from tourism \citep{Fine2019}. Although coral bleaching is currently less of an issue in the northern Red Sea than the south, its potential impact on tourism means studying extreme sea surface temperatures in this region may still be of interest, and we use this example to demonstrate how our model could be employed to investigate the probability of high sea surface temperature events for other locations that may be more at risk of coral bleaching. 

When modelling extreme events through time at a single location, it is usual to decluster the observations into sets of days that correspond to the same event. In practice, this can be achieved using the runs method of \cite{Smith1994}, where clusters of observations containing exceedances above a high threshold $v$ are separated by $r$ consecutive days of non-exceedances. Of interest may be the average length of such clusters, but in studying the effect of high temperatures this may not be the most important factor. In their heatwave application, one alternative proposed by \cite{Winter2016} is to consider the maximum number of consecutive exceedances within a cluster, since these events can have a large impact in practice. We follow this approach, aiming to estimate the number of clusters per year containing a maximum of at least $n$ consecutive exceedances. We first consider events at any single location in the northern Red Sea, and then extend the approach to consider simultaneous exceedances at a set of spatial locations.

\begin{figure}[!htbp]
\begin{center}
\includegraphics[clip, trim=0cm 0.05cm 0cm 0.75cm, width=0.8\textwidth]{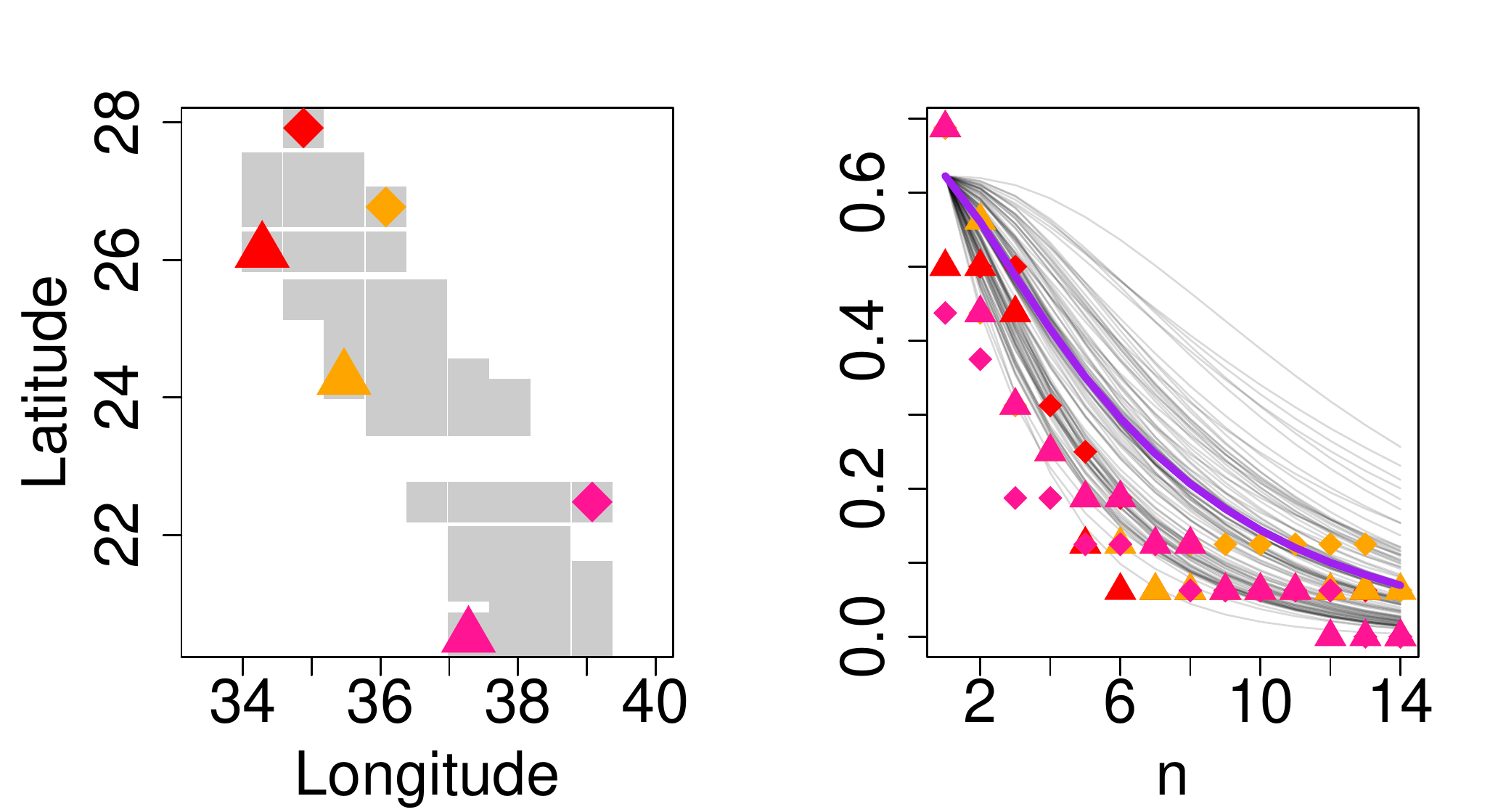}
\caption{Left: six locations where the water depth is less than 150m and coral has the potential to grow. Right: expected number of clusters per year where the maximum number of consecutive exceedances of the 0.961 quantile at a single location is at least $n$, estimated using the fitted model parameters (purple) and bootstrapped parameter estimates (grey); empirical results corresponding to the highlighted locations in the left panel are also shown.}
\label{fig:bleachingResults1}\vspace{1cm}
\includegraphics[clip, trim=0cm 0.05cm 0cm 0.75cm, width=0.8\textwidth]{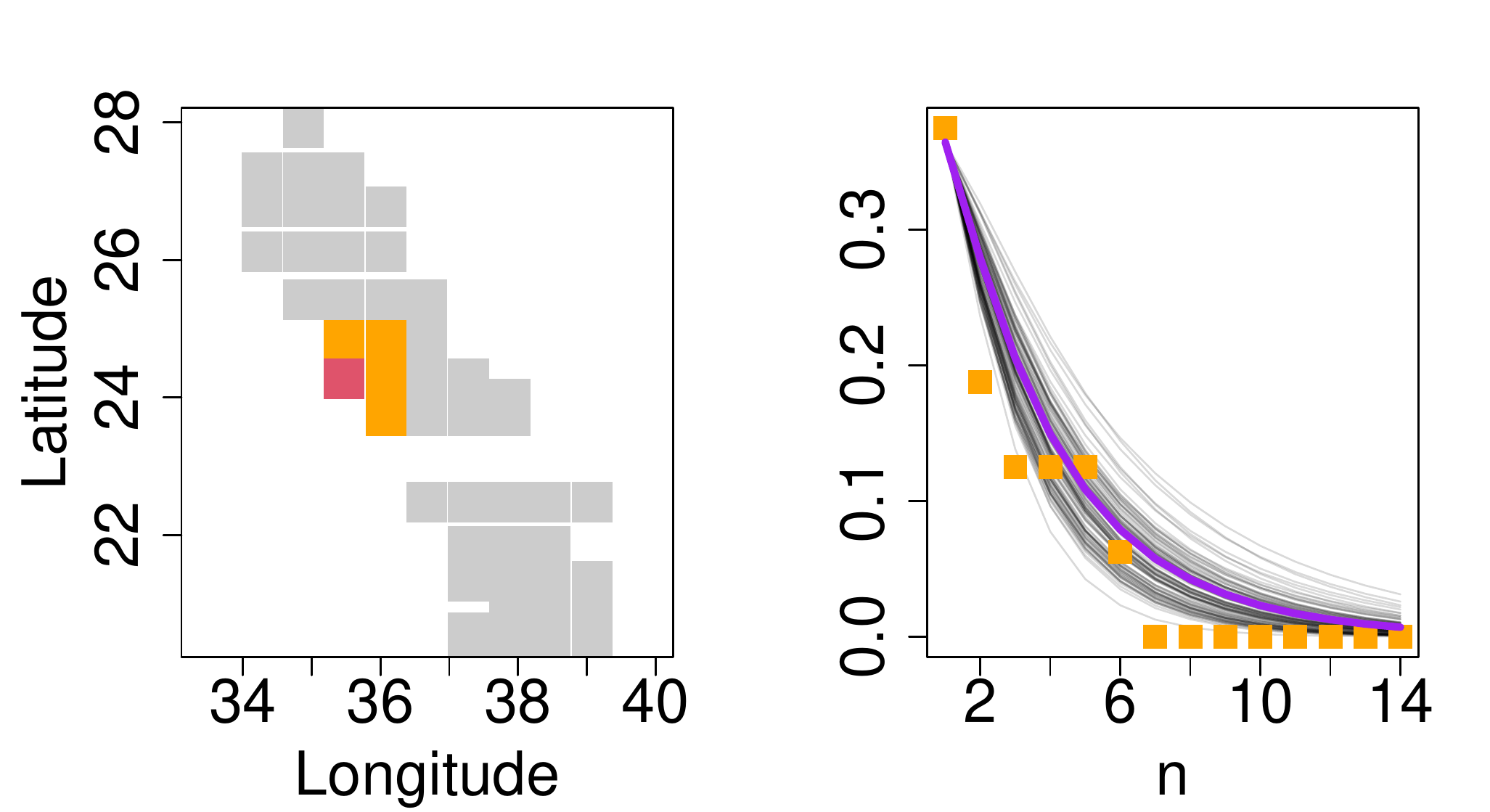}
\caption{Left: one of the shallow-water locations (red) and its neighbours (orange). Right: expected number of clusters per year where the maximum number of consecutive days with exceedances of the 0.961 quantile at all of these locations is at least $n$, estimated using the fitted model parameters (purple) and bootstrapped parameter estimates (grey); empirical estimates are shown in orange.}
\label{fig:bleachingResults2}
\end{center}
\end{figure}

Since corals particularly grow in shallow water, we focus on locations where the water depth is less than 150m \citep{Genevier2019}, shown in the left panel of Figure~\ref{fig:bleachingResults1} using bathymetry data from the \cite{GEBCO2019}. \cite{Winter2016} discuss the importance of \emph{within-cluster} and \emph{over-cluster} results; in our single location setting, the former is the probability that the maximum number of consecutive exceedances in a cluster is at least $n$, while the latter is the expected number of clusters in a single year. Assuming independence of clusters, we can multiply together these two values to obtain our estimate.

To study the expected number of clusters per year, we apply the runs method of \cite{Smith1994}, with parameter $r=10$, at each of the 54 locations, and take empirical values. Due to our assumption of spatial stationarity, we take the average over all these locations to obtain our estimate. We apply this same approach for other values of $r$ in Section~\ref{sup:runsParameter} of the Supplementary Material; $r=10$ corresponds to the value above which the estimated number of clusters stabilises, suggesting independence between clusters. We estimate the distribution of the maximum number of consecutive exceedances in a cluster via simulation, based on our fitted model. To simulate single events, \cite{Smith1997} suggest conditioning on the maximum value within a cluster being above the threshold $v$, and simulating forwards and backwards in time. This can be achieved by applying the simulation method in Section~\ref{subsec:simulation} within a rejection sampling routine, with rejection if the maximum value does not occur at the conditioning site; see \cite{Winter2016}. We obtain 250,000 observations using this technique, and take empirical estimates for the required distribution, ensuring that the conditioning site is within the contributing cluster, and that each cluster begins and ends with $r=10$ non-exceedances of the threshold. This is repeated using each set of bootstrapped parameter estimates. The results are shown in the right panel of Figure~\ref{fig:bleachingResults1}. Here, we also present empirical estimates for the six shallow-water locations, and our model appears to give reasonable results. Further, at four of these six locations, the data does not contain a block of consecutive exceedances longer than 13 days, i.e., our model enables us to consider the expected occurrence of unobserved events.

\cite{Genevier2019} consider events where the sea surface temperature is above some critical level for seven consecutive days. For any single location in the northern Red Sea, we estimate the expected number of clusters with a maximum of at least seven consecutive exceedances to be 0.247 per year. This corresponds to a return period of 4.05 years with a 95\% confidence interval of $(2.28, 9.10)$ years.

Since extreme surface temperatures occurring over larger spatial domains could have more impact on marine life, for one of the locations in Figure~\ref{fig:bleachingResults1} taken as $s_0$, we consider simultaneous extremes at this and its neighbouring sites, which correspond to locations $s$ where $\|s-s_0\|<1$ in the transformed coordinates; these sites are shown in the left panel of Figure~\ref{fig:bleachingResults2}. We now define a cluster as a series of days where all these sites exceed their 0.961 quantile simultaneously, separated by $r=10$ days where this is not the case, and are interested in the maximum number of consecutive, simultaneous exceedances in a cluster. The estimate of the number of clusters is again obtained empirically, but we now only average over those locations with four, five or six neighbours within a distance of 1, to give a reasonable approximation for the location of interest. 

To estimate the distribution of the maximum number of consecutive exceedances within a cluster, we note that our definition of a cluster means we must condition on there being an exceedance at all locations within the cluster, rather than at a single location as in the previous case. We employ a similar rejection sampling technique as previously, simulating forwards and backwards conditioning on $X(w_0)>v$, with the criteria for rejection being that at least one site is below the threshold $v$, and the maximum within the cluster does not occur at the conditioning time.

Results are shown in the right panel of Figure~\ref{fig:bleachingResults2}; we now have no observations of a maximum number of consecutive exceedances greater than six, but are able to extrapolate beyond this using our model. The expected number of clusters per year where all locations exceed their 0.961 quantiles for a maximum of at least seven consecutive days is estimated to be 0.057. This corresponds to a return period of 17.4 years with a 95\% confidence interval of $(9.5, 41.8)$ years.


\section{Discussion}\label{sec:discussion}
In this paper, we presented an approach to modelling extreme events over space and time, by extending the conditional spatial extremes model of \cite{Wadsworth2019} to a spatio-temporal setting. The model is constructed by conditioning on exceedances above a threshold at a single location, with inference carried out via a composite likelihood approach, allowing for contributions from different conditioning sites. We used this approach to model sea surface temperature extremes in the north of the Red Sea, and proposed a range of diagnostic techniques, showing that the data were well described by the model. The resulting model fit was used to demonstrate how one could assess the risk of coral bleaching, by estimating the return period of clusters where sea surface temperatures exceed a high threshold over consecutive days, across one or several spatial locations.

One issue that could be considered further is the differing behaviour observed in the north and south of the Red Sea. We chose to concentrate on modelling surface temperatures only in the north to simplify our approach, but it would but useful to have techniques available to model the full set of locations simultaneously. One way to deal with the non-stationarity in the data may be to allow the model parameters to depend on the spatial conditioning site in some way. We considered allowing the parameters $\lambda_{\textsc{s}}$ or $\kappa_{\textsc{s}}$ to depend on the spatial coordinates of the conditioning site, and separately tried to include water depth as a covariate in the parameter $\lambda_{\textsc{s}}$. Although these attempts did not sufficiently improve our model fit, it is possible that a different covariate could have explained some of the spatial non-stationarity present in the data, and covariate modelling as a technique may be successful in other applications.

A final aspect that could benefit from further attention in the future is that of threshold selection. This is a topic of notorious difficultly within extreme value statistics, with the univariate case seeing a particularly large amount of research; see for instance \cite{Scarrott2012} or \cite{Northrop2017}. The problem lies in choosing a threshold low enough that there is sufficient data to carry out reliable inference, but high enough that the asymptotic assumptions of the model hold. A common technique is to use plots to assess the stability of parameter estimates, or some other measure related to the data, but these plots can be difficult to interpret, and the resulting threshold choice subjective. Our choice to use a fixed quantile in the present paper has the benefit of simplicity, but we acknowledge that, as with all extremal modelling, there may be some sensitivity to this choice.


\section*{Acknowledgements}
This publication is based upon work supported by the King Abdullah University of Science and Technology (KAUST) Office of Sponsored Research (OSR) under Award No.\ OSR-CRG2017-3434.


\bibliography{refs}


\newpage
\appendix
\begin{center}
{\bf\Large Supplementary Material for `Conditional Modelling of Spatio-Temporal Extremes for Red Sea Surface Temperatures'}\\
{\large Emma S.\ Simpson and Jennifer L.\ Wadsworth\\}
Lancaster University
\end{center}

\section{Choice of block-length for inference}\label{sup:differentClusterLength}
\begin{figure}[!htbp]
\begin{center}
\includegraphics[width=0.8\textwidth]{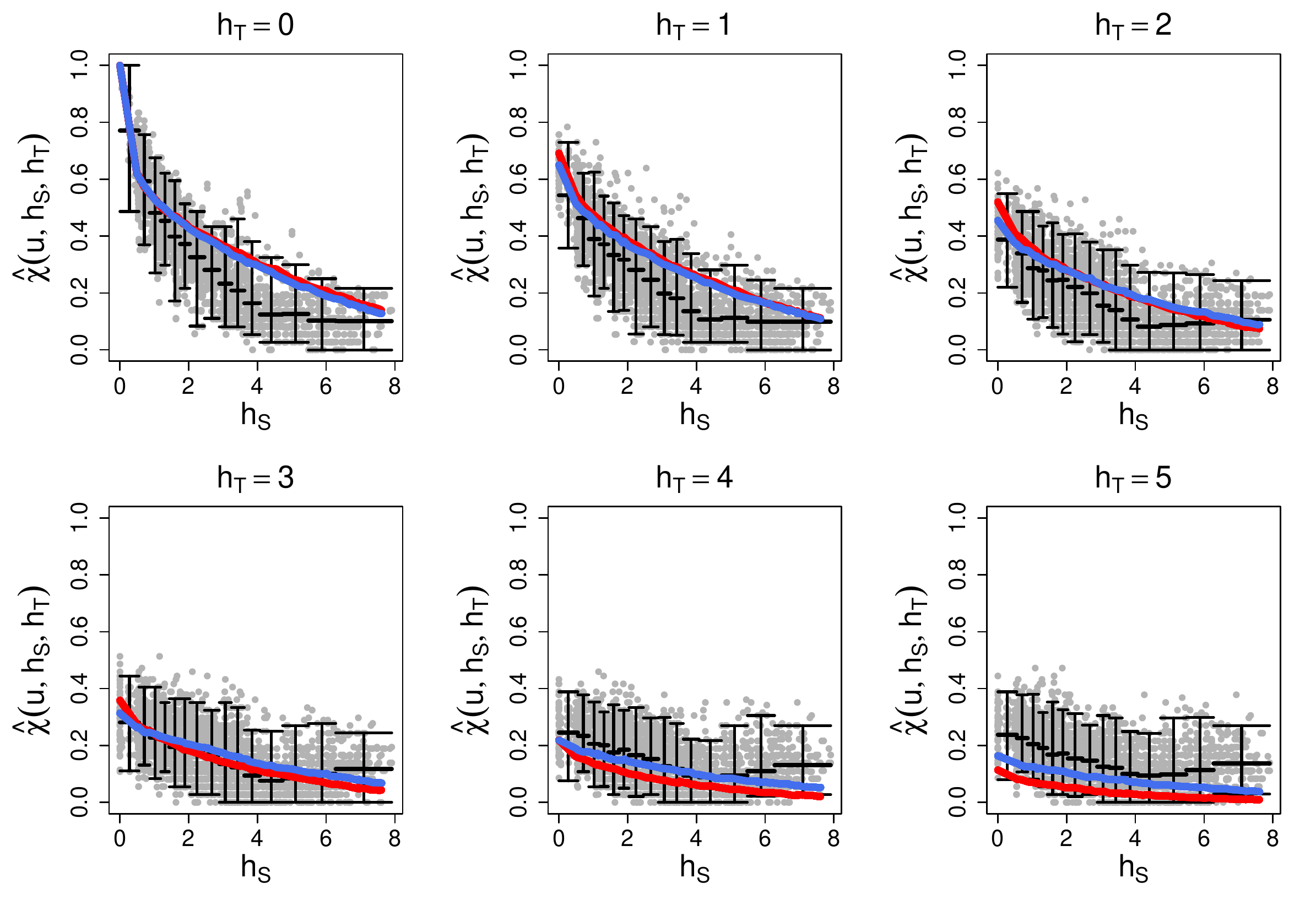}
\caption{Pairwise estimates of $\chi(0.975;h_\textsc{s},h_\textsc{t})$ for the Red Sea data (grey), with the mean and 0.025 and 0.975 quantiles for sections of the distances (black). The solid lines show $\chi(0.975;h_\textsc{s},h_\textsc{t})$ estimates for our fitted models with $m=3$: separable (red); non-separable (blue).}
\label{fig:chi_cluster3_975}
\end{center}
\end{figure}

\begin{figure}
\begin{center}
\includegraphics[width=0.8\textwidth]{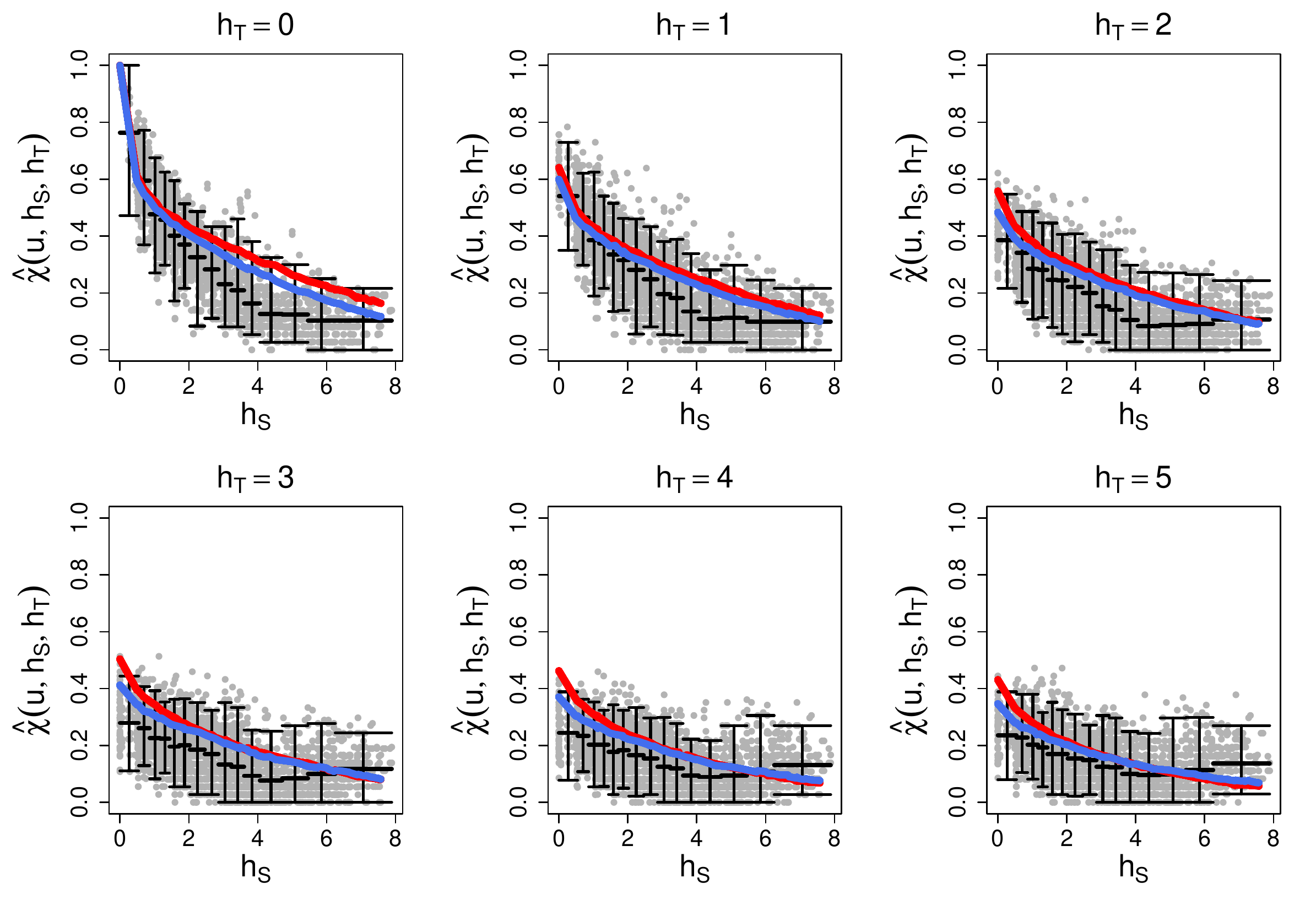}
\caption{Pairwise estimates of $\chi(0.975;h_\textsc{s},h_\textsc{t})$ for the Red Sea data (grey), with the mean and 0.025 and 0.975 quantiles for sections of the distances (black). The solid lines show $\chi(0.975;h_\textsc{s},h_\textsc{t})$ estimates for our fitted models with $m=8$: separable (red); non-separable (blue).}
\label{fig:chi_cluster8_975}
\end{center}
\end{figure}

To create the diagnostic plots in Section~\ref{sec:diagnostics}, we fixed the block-length used for inference at $m=5$. In Section~\ref{subsec:chiSpaceTime}, we compared empirical estimates of $\chi(u;h_\textsc{s},h_\textsc{t})$ for the Red Sea data and simulations from our fitted model, with $a_{w-w_0}(x)$ having both a separable and non-separable form. In Figures~\ref{fig:chi_cluster3_975}~and~\ref{fig:chi_cluster8_975}, we present similar results for $u=0.975$, but taking block-lengths of $m=3$ and $m=8$, respectively. These plots demonstrate that all three block-lengths perform similarly well for time lags less than three, but taking $m=3$ leads to worse results for larger time lags. The results for $m=8$ are very similar to those for our chosen value of $m=5$. That is,  by taking $m=8$, we suffer from increased computation time but do not observe improved results, thus suggesting we have selected an appropriate block-length for our data.



\section{Anisotropy transformation}\label{sup:anisotropy}
\begin{figure}[!htbp]
\begin{center}
\includegraphics[clip, trim=0cm 0.5cm 0cm 1cm, width=0.4\textwidth]{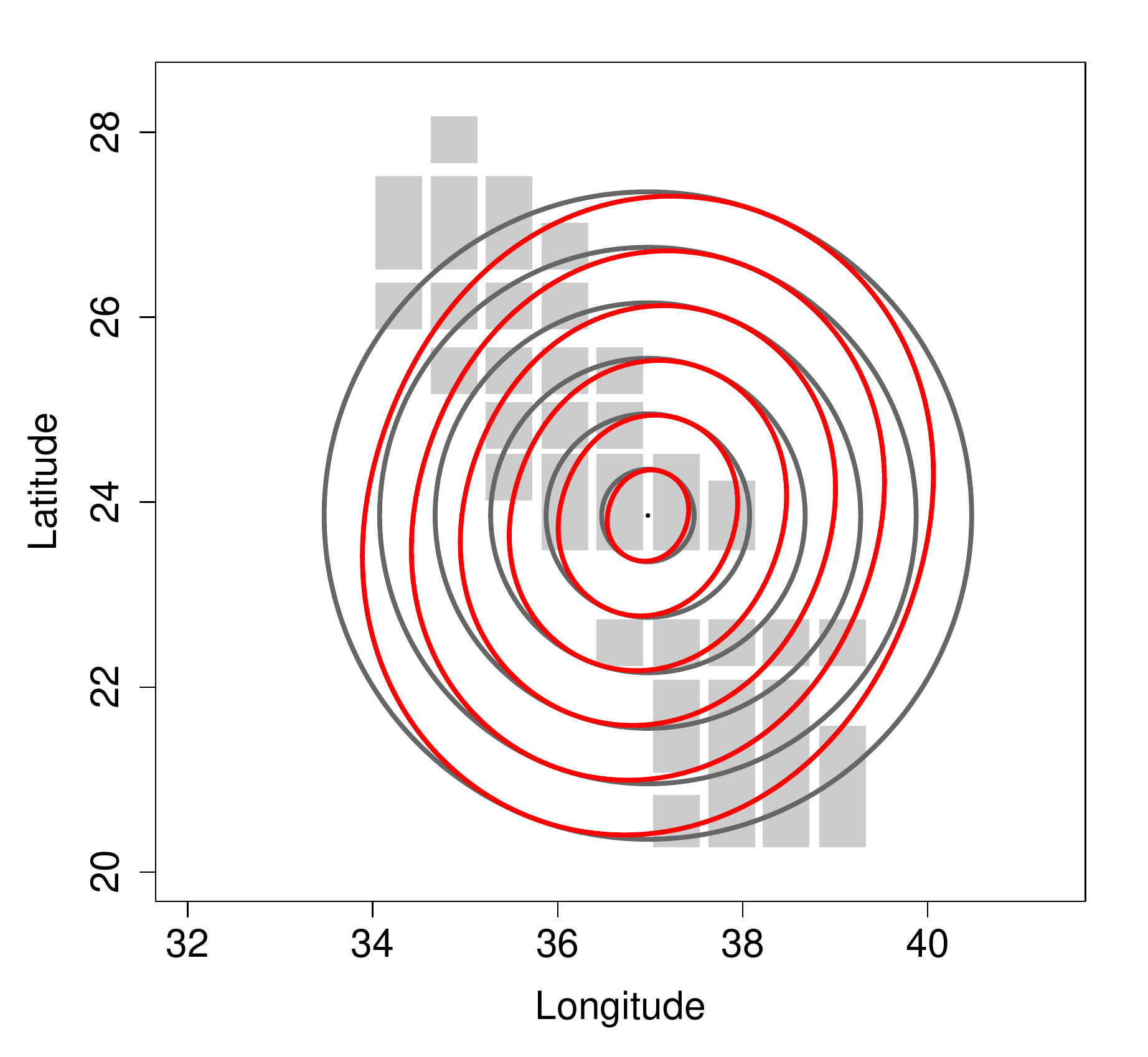}
\caption{Demonstration of geometric anisotropy. Distances under no anisotropy: black; distances under transformation~\eqref{eqn:anisotropy} with $(\theta, L)=(-1.21,1.17)$: red.}
\label{fig:anisotropyFit}
\end{center}
\end{figure}

In Section~\ref{subsec:anisotropy}, we discussed the use of a spatial transformation to account for anisotropy in the observations. This is achieved via equation~\eqref{eqn:anisotropy}, with the transformation controlled by parameters $(\theta,L)$.  For the separable model \eqref{eqn:separable_a} and non-separable model~\eqref{eqn:nonsepAlpha}, respectively, the estimated anisotropy parameters are $(\hat\theta,\hat L)=(-1.22,1.17)$ and $(\hat\theta,\hat L)=(-1.21,1.17)$. We demonstrate the transformation for the latter case in Figure~\ref{fig:anisotropyFit}.

\section{Temporal diagnostic based on $\{Z^0(w)\}$}\label{sup:Zdiagnostic}
\begin{figure}
\begin{center}
\includegraphics[width=0.8\textwidth]{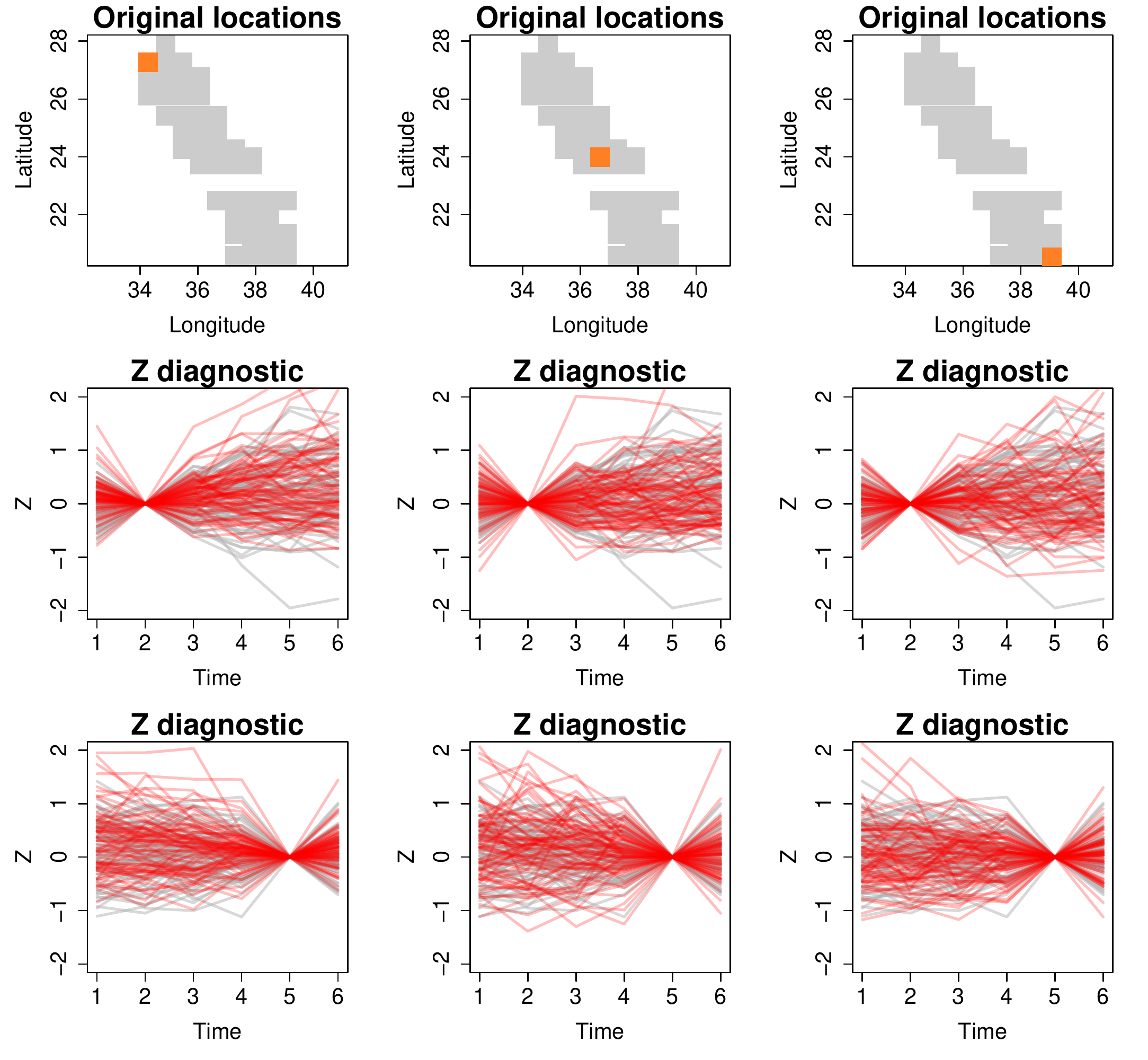}
\caption{Top row: locations of temporal residual process diagnostic (orange); second and third rows: comparison of residuals simulated from the fitted model (grey) and empirical residuals (red) across time, conditioning on extremes at times two and five.}
\label{fig:temporalZdiag_cluster5}
\end{center}
\end{figure}

In Section~\ref{subsec:Zdiagnostic}, we presented a diagnostic to assess our model fit, based on the purely spatial aspect of the residual process $\{Z^0(w)\}$. In Figure~\ref{fig:temporalZdiag_cluster5}, we present similar results to assess the temporal aspect of our model. For three different locations, we observe the behaviour of the residual process through time, taking $t_0$ to be either the second or fifth observation in a block of length six. As in the spatial case, we observe similar behaviour between the empirical residuals and the residuals simulated from the fitted model.


\section{Critical temperature levels for coral bleaching}\label{sup:coralQuantiles}
\begin{figure}[ht]
\begin{center}
\includegraphics[clip, trim=0cm 0cm 0cm 0.5cm, width=0.98\textwidth]{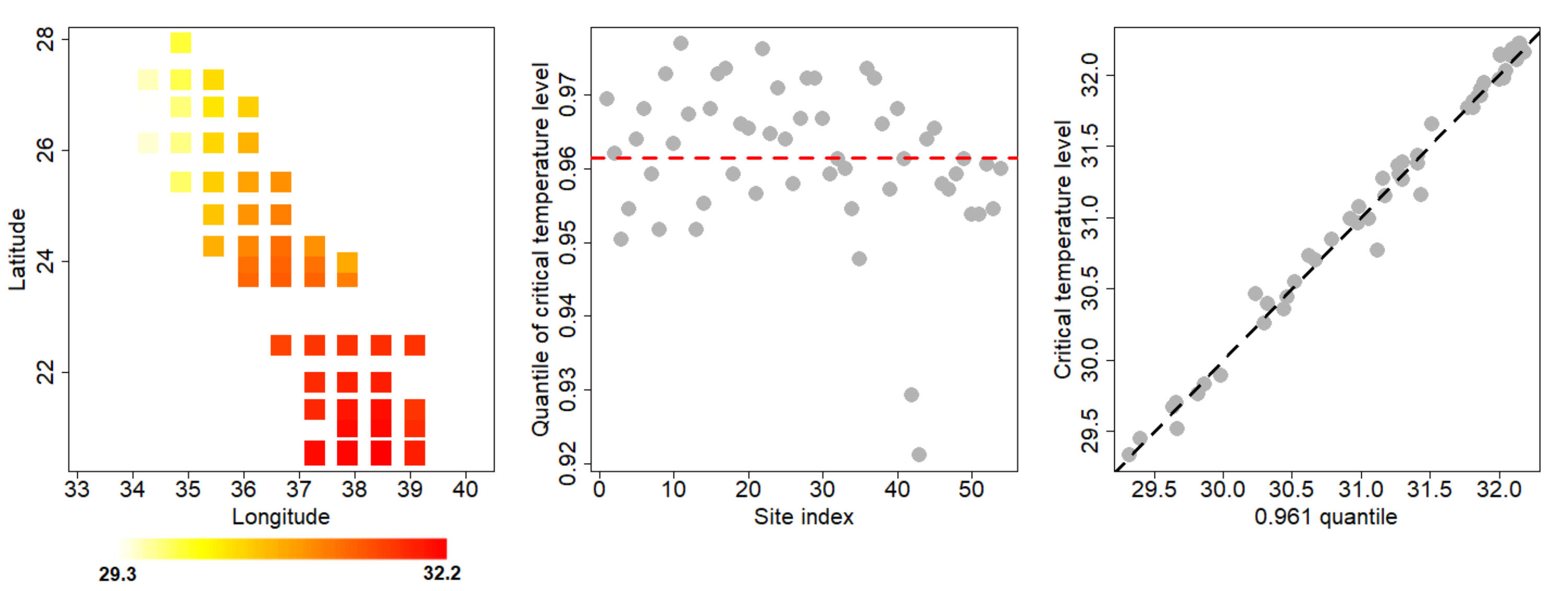}
\caption{Left: one degree more than the average August temperature (\celsius) across the north of the Red Sea. Centre: empirical quantiles corresponding to these critical temperatures. Right: comparison of the critical temperature levels and empirical 0.961 quantiles.}
\label{fig:criticalLevels}
\end{center}
\end{figure}

In Section~\ref{subsec:corals}, we presented results in the context of coral bleaching. To a certain extent, corals are able to adapt to their surroundings, but sea surface temperatures can be problematic if they exceed some critical level. In our analysis, we take these critical levels to be one degree more than the average temperature in the hottest month at each location. For the north of the Red Sea, this hottest month is August; we show the critical sea surface temperature levels in the left panel of Figure~\ref{fig:criticalLevels}. 

Each of these critical values can be associated with an empirical quantile of the temperatures at its corresponding location, as shown in the centre panel of Figure~\ref{fig:criticalLevels}. On average, this corresponds to the 0.961 quantile of the temperatures across the summer months; we use this quantile when working on the Laplace scale in Section~\ref{subsec:corals}. Finally, in the right panel of Figure~\ref{fig:criticalLevels}, we plot the critical temperature values against the empirical 0.961 quantile for each location. These demonstrate reasonable agreement, suggesting that the critical temperature levels are well represented by this quantile level.


\section{Choice of declustering parameter}\label{sup:runsParameter}
\begin{figure}[!htbp]
\begin{center}
\includegraphics[width=0.5\textwidth]{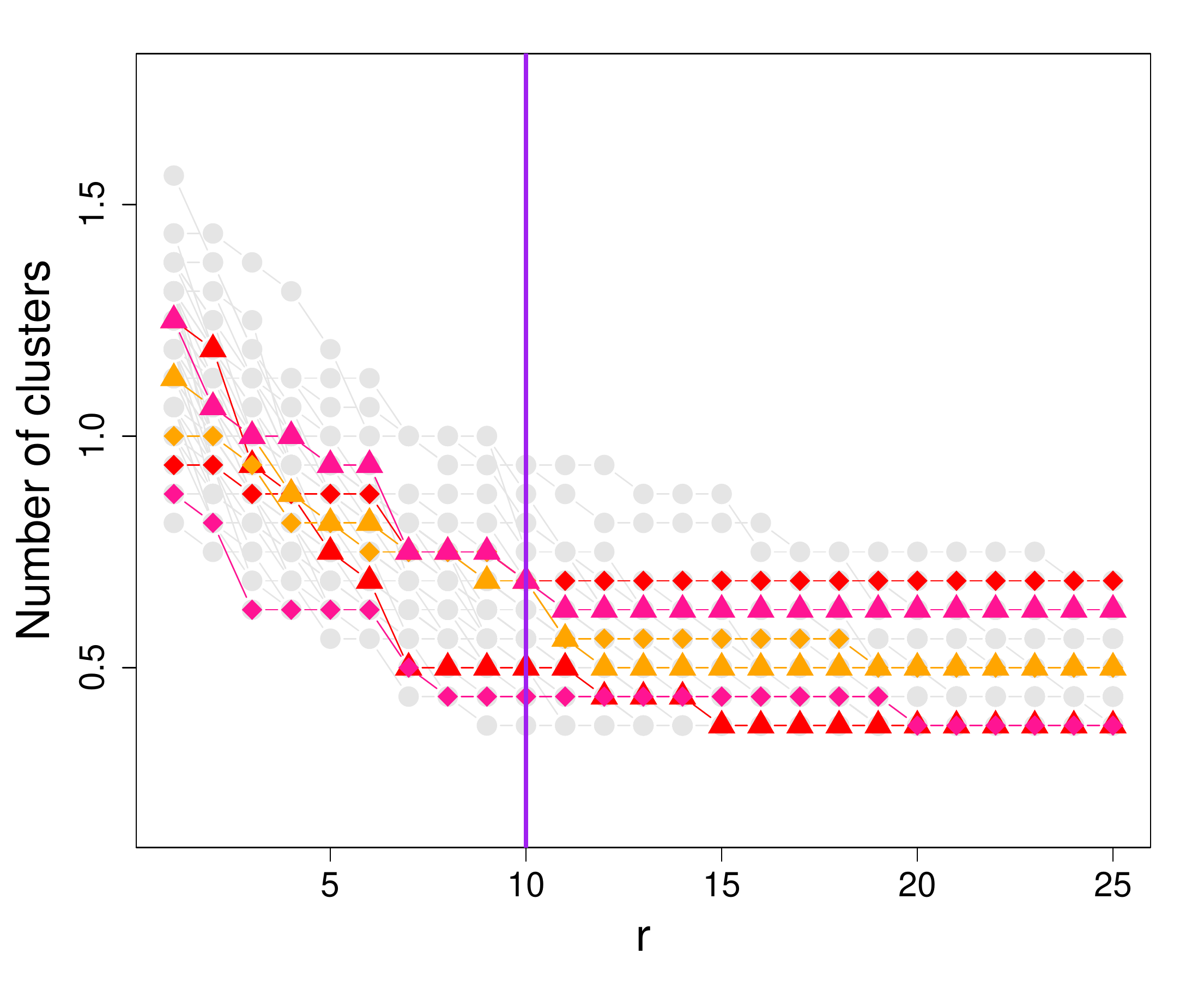}
\caption{Grey: estimated number of clusters per year for each of the 54 locations in the north of the Red Sea, with declustering parameter $r\in\{1,\dots,25\}$. The highlighted lines correspond to the shallow-water locations in the left panel of Figure~\ref{fig:bleachingResults1}.}
\label{fig:clusterNumbers}
\end{center}
\end{figure}

In Section~\ref{subsec:corals}, we applied the runs method of \cite{Smith1994} to temporally decluster extreme events at individual locations. In particular, we fixed the parameter of the method to $r=10$, and to estimate the expected number of clusters per year, averaged empirical results from across all 54 locations. In Figure~\ref{fig:clusterNumbers}, we present results for each location using different values of $r$, demonstrating that the estimates become stable above our chosen parameter value.

\end{document}